\documentclass{ametsocV6.1}
\draftfalse

\usepackage{amsmath,amsfonts,bm}

\def\eqref#1{\ref{#1}}

\def\1{\bm{1}}

\DeclareMathAlphabet{\mathsfit}{\encodingdefault}{\sfdefault}{m}{sl}
\SetMathAlphabet{\mathsfit}{bold}{\encodingdefault}{\sfdefault}{bx}{n}

\newcommand{\KL}{D_{\mathrm{KL}}}

\usepackage{url} 
\usepackage{soul}
\usepackage[utf8]{inputenc} 
\usepackage[T1]{fontenc}    
\usepackage{url}            
\usepackage{booktabs}       
\usepackage{multirow}
\usepackage{nicefrac}       
\usepackage{tikz}
\usepackage{svg}

\usepackage{array}
\usepackage{graphicx}
\usepackage{paralist}
\usepackage{float}
\usepackage[stable]{footmisc}
\usepackage{tikz}
\usepackage{enumitem}
\usepackage{amsmath}
\usepackage{amssymb}
\usepackage{mathtools}

\usepackage{amsthm}
\usepackage{algorithm}
\usepackage{algorithmic}
\usepackage{amsfonts}       
\usepackage{relsize} 

\usepackage[percent]{overpic} 
\usepackage{xcolor}     
\usepackage{tikz}       
\usetikzlibrary{arrows.meta}

\usepackage[most]{tcolorbox}
\usepackage{makecell}

\usepackage{siunitx}
\theoremstyle{plain}

\theoremstyle{definition}

\theoremstyle{remark}

\newcommand{\EE}{\mathbb{E}}

\newcommand{\diff}{\mathrm{d}}

\newcommand{\SSS}{\mathcal{S}}
\newcommand{\TT}{\mathcal{T}}
\newcommand{\TTx}{\TT_{\text{x}}}
\newcommand{\TTy}{\TT_{\text{y}}}

\newcommand{\gen}[1]{#1^{\text{gen}}}

\DeclareMathOperator*{\SW}{SW}

\makeatletter
\@ifpackageloaded{lineno}{%
  \let\linenumbers\relax

}{}
\makeatother

\title{Reconstructing the Aerosol State from Partial Observations with Generative Modeling}

\authors{
	Ehsan Saleh,\aff{a,d,f,$*$}\thanks{$^{*}$Work done while at the University of Illinois Urbana-Champaign},
	Saba Ghaffari,\aff{a,d}
	Jeffrey H. Curtis,\aff{c}
	Lekha Patel,\aff{e}
	Peter A. Bosler,\aff{e}
	Nicole Riemer,\aff{c}\correspondingauthor{Nicole Riemer, nriemer@illinois.edu}
	and Matthew West\aff{b}
}

\affiliation{\aff{a}{Department of Computer Science,\\University of Illinois Urbana-Champaign, Urbana, IL, USA}\\
\aff{b}{Department of Mechanical Science and Engineering,\\University of Illinois Urbana-Champaign, Urbana, IL, USA}\\
\aff{c}{Department of Climate, Meteorology and Atmospheric Sciences, University of Illinois Urbana-Champaign, Urbana, IL, USA}\\
\aff{d}{National Center for Supercomputing Applications,\\University of Illinois Urbana-Champaign, Urbana, IL, USA}\\
\aff{e}{Center for Computing Research,\\Sandia National Laboratories, Albuquerque, NM, USA}\\
\aff{f}{Now at: Amazon, Sunnyvale, CA, USA}}

\abstract{Key aerosol properties that shape climate—such as CCN
  activity, scattering and absorption, and ice nucleation
  efficiency—are difficult to infer from measurements that typically
  capture only a part of the aerosol state.  We develop a conditional
  generative framework that maps a label (a vector of partial
  observations) to an ensemble of plausible aerosol states and
  propagates these to diagnostics, yielding mean estimates with
  confidence intervals. Using synthetic data, we evaluate two label
  configurations: a low-dimensional setup with limited number
  distribution and bulk-composition information, and a
  high-dimensional setup including complete number and total mass
  distributions plus species bulk masses. Generated samples maintain
  strong label compliance, and higher-dimensional labels markedly
  reduce variability.  CCN activity and volume scattering are well
  constrained even under the low-dimensional setup, whereas dust- and
  BC-sensitive diagnostics (frozen fraction, absorption) benefit
  substantially from the additional information in the
  high-dimensional case. This framework clarifies which observational
  inputs most effectively constrain different diagnostics and
  demonstrates how generative machine learning can provide
  uncertainty-aware estimates from incomplete aerosol information.}

\begin{document}

\maketitle
\onecolumn
\makeatletter\@ifpackageloaded{lineno}{\nolinenumbers}{}\makeatother
\section{Introduction}
Atmospheric aerosols—suspensions of small liquid or solid
particles—play a central role in both climate and air 
quality~\citep{Seinfeld2016,Poeschl2005}. They scatter and absorb solar
radiation~\citep{Haywood2000}, act as seeds for cloud droplet and ice
crystal formation~\citep{Andreae2008}, and affect human health when
inhaled~\citep{Pope2006}. Despite their importance, aerosols remain
one of the largest sources of uncertainty in climate projections
~\citep{IPCC2021}, in part because their properties vary strongly in
space and time and depend sensitively on particle size, composition,
and mixing state~\citep{McFiggans2006,Riemer2019}.

A convenient way to describe this complexity is through the aerosol
state: the joint distribution of particle sizes, chemical
compositions, and other attributes across the population~\citep{Riemer2024}. 
The aerosol state provides the foundation for diagnosing climate-relevant 
properties such as cloud condensation nuclei (CCN) activity, optical 
scattering and absorption, and ice-nucleating particle (INP) concentrations. 
If the aerosol state were fully known, these emergent properties could be 
computed directly with relatively little ambiguity.

In practice, however, no single instrument can capture the full
aerosol state~\citep{Kulkarni2011,Riemer2019,Kahn2023}. Instead, each
probes a restricted aspect of the population. Mobility spectrometers
such as the Scanning Mobility Particle Sizer (SMPS,~\citet{Wang1990})
resolve submicron number distributions, while aerodynamic particle
sizers (APS,~\citet{Kulkarni2011}) extend into larger
diameters. Chemical composition is even harder to constrain:
instruments like the Aerosol Chemical Speciation Monitor (ACSM,~\citet{Ng2011}) 
provide only bulk averages across all particles, the
Aerosol Mass Spectrometer (AMS,~\citet{Jayne2000}) offers
size-resolved but chemically selective information, and
single-particle techniques give detailed snapshots that are difficult
to quantify at the population level. As a result, constructing a more
complete view of the aerosol state requires combining multiple
measurements, a resource-intensive effort typically undertaken only
during specialized field campaigns involving large teams of
investigators~\citep{Kahn2023}.

Generative machine learning provides a natural way to address this
challenge. Conditional generative models can map from incomplete
inputs to ensembles of plausible aerosol states, enabling uncertainty
quantification and robust estimation of derived diagnostics. Among
available approaches, conditional variational autoencoders (CVAEs;
~\citet{sohn2015learning}) offer a pragmatic balance of stable
training, efficient inference, and well-calibrated uncertainty
compared to alternatives such as GANs~\citep{Mirza2014},
flows~\citep{Rezende2015}, or diffusion
models~\citep{Dhariwal2021}. CAEs are also flexible with respect to
measurement definitions and diagnostic operators, making it readily
extensible to diverse observational setups. However, each definition
of the measurement input requires its own trained CVAE; if the
measurement configuration changes the model must be retrained to learn
the corresponding mapping.

In this study, we investigate whether partial observations can be
combined with machine learning to recover the aerosol state
sufficiently well to enable reliable estimation of climate-relevant
diagnostics such as CCN activity, optical properties, and INP
concentrations. To this end, we use synthetic data from the detailed
aerosol model PartMC-MOSAIC~\citep{Riemer2009,Zaveri2008}, aggregated
into binned size–composition distributions. This data provides a
controlled setting to train a CVAE that reconstructs more complete
aerosol states from partial inputs, allowing us to evaluate its
ability to recover the key diagnostics of interest.

We test our approach using two levels of partial observations as labels. In
the low-dimensional label setup, we combine number distributions over a restricted
range—similar to those obtained from an SMPS—with bulk mass concentrations
analogous to an ACSM. In the high-dimensional label setup, we provide a more
detailed view consisting of the complete number distribution, the full total
mass distribution, and bulk masses of all species. Comparing the two cases
allows us to assess how the richness of the observations influences the
reconstruction of the aerosol state and the reliability of derived diagnostics.
In our proof-of-concept study, both training and inference are performed with
synthetic data generated by a detailed aerosol model. While no real measurements
are used here, our approach establishes the framework and is designed to inform
future integration of diverse in situ observations.

Classical inversion and data-assimilation techniques also aim to
recover aspects of the aerosol state from partial observations. For
example, GRASP-based retrievals invert multi-angle and multi-spectral
radiances to estimate columnar aerosol composition and microphysical
properties~\citep{li2019retrieval,zhang2020improved,dubovik2021comprehensive},
while statistical inversions reconstruct number distributions from
discretized
counters~\citep{mcguffin2020novel,voutilainen2001statistical}, and
ensemble or variational data assimilation fuses satellite AOD (aerosol
optical depth) with chemical transport models to obtain 3-D aerosol
fields~\citep{pagowski2012experiments,Benedetti2009,Randles2017,Choi2020}.
These inversion and data-assimilation approaches rely on explicit
forward models of aerosol physics and optics, and each new measurement
type typically requires a bespoke forward–inverse formulation. In
contrast, our approach replaces the explicit physical forward model
with a data-driven generative model that learns the mapping from
partial observations to ensembles of plausible aerosol states. Our
method can flexibly incorporate new combinations of inputs and
propagate their uncertainties to climate-relevant diagnostics. Unlike
traditional inversions, no hand-crafted physics operator is required
for each measurement type, making the method more extensible across
diverse observational setups.

To date, no study has systematically applied generative machine
learning to infer the aerosol state from partial in situ-like
measurements in a way that directly targets climate-relevant
diagnostics. Our work addresses this gap. Our contributions are
fourfold: (1) we introduce a conditional generative framework for
estimating aerosol states and derived diagnostics from limited
measurement inputs, (2) we propose a novel Wasserstein regularization
loss to improve label compliance, (3) we evaluate its performance
across two realistic observational setups (low- and high-dimensional
labels), and (4) we demonstrate its ability to reproduce CCN activity,
optical properties, and INP concentrations with quantified
uncertainty. In this proof-of-concept study, both training and
inference rely entirely on synthetic data, which provides a controlled
setting to establish and test the method. More broadly, our study
highlights how generative ML approaches can complement traditional
physics-based methods in the Earth sciences by transforming
incomplete, heterogeneous observations into actionable information for
climate and air-quality research.

\section{Conceptual Workflow for Aerosol State Reconstruction}\label{sec:02condgen}


\begin{figure*}
	\centering
	\includegraphics[width=0.98\linewidth]{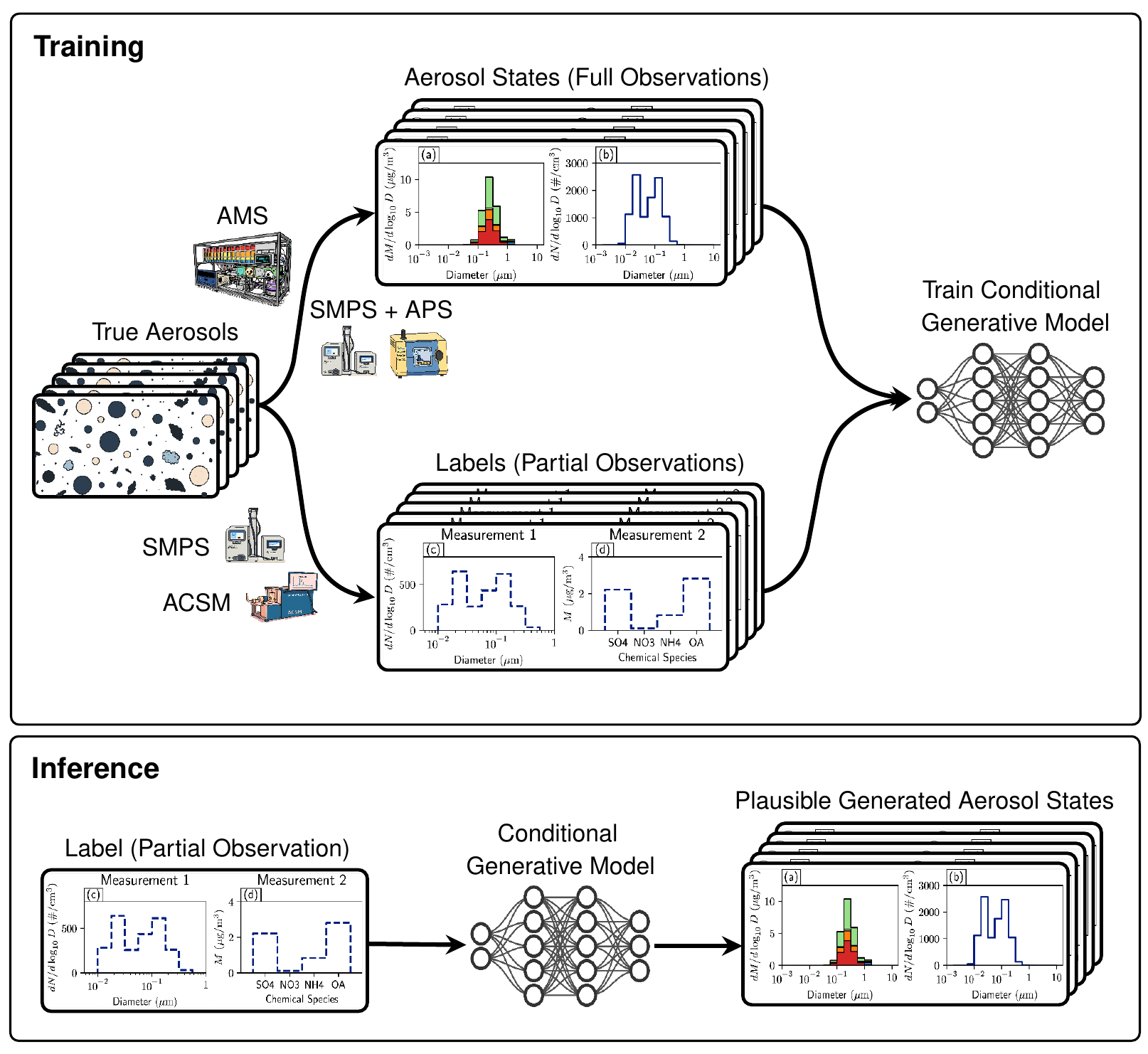}
	\caption{Conceptual overview of the training and inference procedures. True
          aerosol populations (top left) can be characterized in principle by
          full observations of the aerosol state, such as size- and
          composition-resolved distributions measured by instruments like the
          Aerosol Mass Spectrometer (AMS) together with a unified number
          distribution measured by a combination of the Scanning Mobility
          Particle Sizer (SMPS) and an Aerodynamic Particle Sizer (APS). In
          practice, however, such comprehensive measurements are rarely
          available. Instead, partial observations (bottom left) are more
          common, such as a truncated number distribution from the SMPS and
          limited species bulk masses from the Aerosol Chemical Speciation
          Monitor (ACSM). These partial inputs are paired with full aerosol
          states during training to teach a conditional generative model (CVAE)
          the mapping from limited observations to plausible complete states.
          During inference, only a single partial observation is needed; from
          this the trained model generates an ensemble of consistent aerosol
          states.}
	\label{fig:01vision}
\end{figure*}

Figure~\ref{fig:01vision} outlines the overall workflow. The figure
illustrates cartoon versions of common aerosol instruments to
highlight the kinds of measurements we aim to emulate (e.g., SMPS,
ACSM, AMS). In this proof-of-concept study, however, both training and
inference are performed with synthetic data generated by a detailed
aerosol model rather than real observations. The method has two
stages: training and inference.

A conditional variational autoencoder (CVAE) is a type of generative model that
learns to represent complex data in a lower-dimensional latent space. During
training, an encoder maps states and labels into this latent space, and a
decoder reconstructs the corresponding full state from both the latent
representation and the label. Once trained, the model can generate many
plausible states consistent with a given label by sampling different
points in latent space. This structure makes CVAEs well suited for problems like
ours, where partial observations must be mapped to uncertain but physically
plausible full states.

In this context, we define a reconstructed state as plausible if it is
statistically consistent with the distribution of aerosol states
encountered during training and yields physically reasonable
diagnostic quantities. This means that generated states must respect
basic constraints such as non-negative species masses and realistic
size distributions, and they typically fall within observed
climatological ranges for key diagnostics such as CCN, extinction, or
absorption. Importantly, the stochastic sampling of the CVAE does not
produce arbitrary “hallucinations,” but instead expresses epistemic
uncertainty in underconstrained regions by drawing on patterns learned
during training. As such, the variability across generated states
should be interpreted as the range of plausible completions consistent
with the available labels, rather than a deterministic prediction.

Training a CVAE requires many matched pairs of states and labels. In
our context, the \emph{state} represents a complete-as-possible observation
of the aerosol population, including the particle number distribution
and size-resolved mass distributions for individual species. In
principle, such a state could be obtained from a comprehensive suite
of instruments combining size and composition information, though
doing so is logistically demanding and rarely feasible. In our study,
we use synthetic states generated by a detailed aerosol model (see
Section~\ref{sec:03data}), aggregated into binned number and speciated mass
distributions. Each state is represented as a 320-dimensional vector.

The \emph{label} represents a partial observation of the aerosol state, akin to
what can be obtained more routinely from a limited set of instruments. We
consider two cases: (1) a low-dimensional label setup, with 11 dimensions
corresponding to the number distribution over a restricted size range (similar
to an SMPS) and bulk species concentrations (similar to an ACSM); and (2) a
high-dimensional label setup, with 55 dimensions including the full number
distribution, the full total mass distribution, and bulk masses of all species.
By training on many state-label pairs, the CVAE learns to generate full aerosol
states that are statistically consistent with the given label.
Because the definition of the label differs between these two setups, we train a
separate CVAE for each. In general, any change to the observational definition
requires retraining the model so that it learns the mapping between that
specific type of partial observation and the full aerosol state.

For this proof-of-concept study, inference also uses synthetic data:
the CVAE is applied to new test labels drawn from the synthetic
dataset. Given one such label, the CVAE generates an ensemble of
plausible aerosol states consistent with that input. From each
generated state, we calculate climate-relevant diagnostics, including
CCN activity, optical scattering and absorption coefficients, and
frozen-fraction spectra. Averaging across the ensemble provides mean
predictions, while the spread across samples quantifies uncertainty
arising from the incomplete observational constraints. In future work,
this framework could be applied directly to real measurements, but
here we focus on synthetic data to establish and evaluate the method.

\section{Synthetic Data, State Representation, and Training Protocol} \label{sec:03data}

\subsection{Scenario library and state representation}
We use the aerosol scenario library introduced 
by~\citet{zheng2021estimating}, generated with the particle-resolved
PartMC--MOSAIC framework~\citep{Riemer2009,Zaveri2008}. PartMC
explicitly tracks the composition and size of thousands of
computational particles to resolve mixing state, while MOSAIC provides
coupled gas- and aerosol-phase chemistry and thermodynamics.  Together
they represent emissions, coagulation, dilution, and gas--aerosol
partitioning across diverse conditions.  The library contains 1000
scenarios with 25 hourly snapshots each and includes 15 chemical
species spanning major primary and secondary inorganic and organic
aerosol components.

For this study, we do not use the full particle-resolved output. Instead, each
sample is represented as $x=(m,n)$, where $m$ denotes the speciated mass
distribution and $n$ the number distribution across diameter.  We discretize
diameter into 20 logarithmically spaced bins from $1\,{\rm nm}$ to $10\,{\rm \mu
m}$. For species $a$ and size bin $b$, $m_{a,b}$ is the discrete speciated mass
size distribution, and $n_b$ is the discrete particle number size distribution.
In other words, $m_{a,b}$ and $n_b$ describe how aerosol mass and number are
distributed across particle sizes. This representation yields a 320-dimensional
state vector (20 bins $\times$ 15 species for mass, plus 20 bins for number),
against which model performance is evaluated.  

\subsection{Label definitions and diagnostics}\label{sec:033aeromeasures}
From these states, we construct partial ``observation'' vectors that
mimic what could be obtained by more limited instrument combinations.
We consider two synthetic cases (see Figure~\ref{fig:01vision} for
schematic instrument analogs):

\begin{itemize}
    \item \textbf{Low-dimensional label setup:} 11 dimensions,
      consisting of (1) a truncated number distribution, consisting of
      7-binscovering 10--560\,nm, analogous to the range measured by a
      Scanning Mobility Particle Sizer (SMPS), and (2) limited species
      bulk masses, consisting of the bulk masses for 4
      species---sulfate, nitrate, ammonium, and total organics (the
      latter combining several primary and secondary organic
      species)---analogous to an Aerosol Chemical Speciation Monitor
      (ACSM). Importantly, this setup does not include black carbon or
      dust, species that are also not quantified by ACSM, and our
      results highlight the impact of this omission on absorption and
      ice-nucleation diagnostics. (Note: ACSM can also report
      chlorine, but typically only from gas-phase partitioning; our
      scenario library does not include chlorine chemistry, so this
      channel was omitted.)
    \item \textbf{High-dimensional label setup:} 55 dimensions,
      consisting of (1) the full 20-bin number distribution, (2) the
      full 20-bin total mass distribution, and (3) bulk masses for
      all 15 species, including black carbon and dust. This richer
      representation provides direct constraints on the species most
      critical for absorption and ice nucleation.
\end{itemize}

Each label definition requires training a separate CVAE to learn the mapping
from that particular label to the full aerosol state.

From the states---either true or generated---we compute four climate-relevant
diagnostic spectra: (1) CCN activation 
spectra~\citep{ching2017metrics,fierce2016black,wang2010importance}; (2) volume
scattering coefficients $\beta_s$, computed with the Toon--Ackerman algorithm
for Mie calculations~\citep{toon1981algorithms,geiss2024neuralmie}; (3) volume
absorption coefficients $\beta_a$, computed analogously; and (4) frozen fraction
spectra, treating dust and black carbon as potential ice-nucleating 
particles~\citep{niemand2012particle,schill2020contribution}.

\subsection{Training and evaluation protocol}
To avoid temporal leakage, we split entire simulation trajectories
into training and testing sets using an 80--20 partition.  We repeat
this procedure across 10 random seeds, train a separate model per
split, and report statistics as averages over the 10 runs to reduce
variance and improve robustness. In this study we adopt the fully
synthetic setting, where both training and inference use simulated
states and labels, but the framework is more general and could also be
applied in hybrid or fully measurement-based configurations.

\section{Model}\label{sec:04model}

Conditional variational autoencoders (CVAEs) are generative models that condition both the encoder and decoder on a label vector to learn distributions over states consistent with observations.
To generate new samples from labels alone, the encoder is bypassed and multiple latent draws are sampled from a standard normal distribution and combined with the input labels before decoding.
Each random latent coupled with the label produces a distinct realization, yielding diverse aerosol states while preserving the conditional structure.
With adequate training, the decoder should produce states whose computed labels closely comply with the input labels, supporting reliable downstream analysis.
Section~\ref{sec:041cvae} summarizes the conventional CVAE setup for reconstruction and conditional generation.
Section~\ref{sec:042swcvae} introduces our Wasserstein-regularized CVAE for improved label compliance, which is used for all results in the main paper (see the appendix for a comparison of traditional CVAE results to ours).

\subsection{Conditional Variational Auto-Encoding (CVAE)}\label{sec:041cvae}


\begin{figure*}
	\centering
	\includegraphics[width=0.99\linewidth]{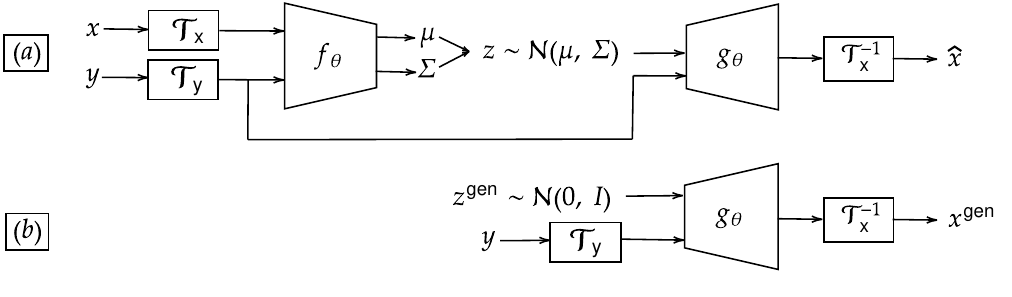}
	\caption{The model structure. Here $f_\theta$ is the encoder network, $g_\theta$ is the decoder network, $\TTx$ and $\TTy$ are preprocessing transformations, and $h$ is the label computation function. (a) The sample reconstruction pipeline takes a state $x$ and its label $y$ and reconstructs the approximate state $\hat{x}$. (b) The conditional generation pipeline takes a randomly-sampled latent vector $\gen{z}$ and label $y$ and generates a plausible state $\gen{x}$.}
	\label{fig:02schematic}
\end{figure*}

Our base generative model is a conditional VAE (Figure~\ref{fig:02schematic}) that conditions on the label $y$ while learning a latent $z$ representation of the input sample $x$.
We first preprocess $x$ and $y$ using $\TTx$ and $\TTy$, following the conventions established in~\citet{saleh2025generative}.
The encoder $f_\theta$ maps the preprocessed inputs to a diagonal-covariance Gaussian with parameters $\mu$ and $\Sigma$, defining the variational posterior over the latent space as
\begin{equation}\label{eq:04encmu}
(\mu,\, \Sigma) = f_\theta\big(\TTx(x),\, \TTy(y)\big).
\end{equation}
The decoder $g_\theta$ combines a latent draw $z$ with the preprocessed label to approximate the preprocessed state, and we apply the inverse transform to recover the reconstruction by
\begin{equation}\label{eq:04xhat}
\hat{x} = \TTx^{-1}\big(g_{\theta}(z,\, \TTy(y))\big).
\end{equation}
Reconstructed diagnostic variables are then computed from $\hat{x}=(\hat{m},\, \hat{n})$.

The model can be trained by minimizing the reconstruction and variational losses:
\begin{equation}\label{eq:04trnloss}
	\mathcal{L} = \EE\Big[\|x - \hat{x}\|_2^2\Big] + w_{\text{KL}} \EE\Big[\KL\big(\mathcal{N}(\mu, \Sigma)\parallel\mathcal{N}(0, I)\big)\Big].
\end{equation}
Here, $\KL$ denotes the KL-divergence, $w_{\text{KL}}$ is a weighting term, and the expectations are replaced with an empirical average over a mini-batch of training samples for stochastic gradient descent optimization.

To generate new samples from measurements alone, we bypass the encoder, draw $\gen{z}\sim\mathcal{N}(0,I)$, decode it jointly with the preprocessed label to obtain a realization, and invert preprocessing to the original state, giving
\begin{equation}\label{eq:04xgen}
\gen{x} = \TTx^{-1}\big(g_{\theta}(\gen{z},\,\TTy(y))\big).
\end{equation}
The randomness in $\gen{z}$ induces diversity across generated states while preserving conditional structure from the input label.

For each generated sample, we compute the label deterministically via the label computation operator, i.e., $y=h(x)$ and $\gen{y}=h(\gen{x})$ (see Section~\ref{sec:033aeromeasures}). It is crucial that $\gen{y}$ closely agrees with the input $y$; otherwise the generated states are not consistent with the input labels, motivating the compliance treatment in Section~\ref{sec:042swcvae}.

\subsection{Wasserstein-Regularized CVAE}\label{sec:042swcvae}

Ensuring that the decoder produces states consistent with the input labels is
essential: a random conditional draw $\gen{x}$ must yield a computed label
$\gen{y}$ that closely matches the input label $y$, otherwise the generated
states are not conditionally valid and downstream diagnostics would be
misleading~\citep{sohn2015learning,bond2021deep}. Appendix~\ref{sec:a02catlbl}
illustrates that conventional CVAEs can violate this requirement, motivating an
explicit notion of \textit{label compliance} that we quantify using the average
relative error. That is, given an input label $y$, and $\gen{y}_i$ being the
computed label from the generated sample $\gen{x}_i$, we define the compliance
error for this input label as
\begin{equation}\label{eq:05comperr}
	\text{Compliance Error} = \frac{1}{N} \sum_{i=1}^{N} \frac{\|y - \gen{y}_i\|_2}{\|y\|_2 + \|\gen{y}_i\|_2}.
\end{equation}

Conventional CVAEs prioritize reconstruction from latent-label inputs, and indeed can reconstruct accurately even when relying primarily on the latent representation, as noted in~\citet{saleh2025generative}. While reconstructions that use the encoded $z$ (cf. Equation~\eqref{eq:04xhat}) tend to exhibit low compliance error, generation with random latent draws $\gen{z}$ can increase the mismatch between $\gen{y}$ and $y$, indicating leakage of label information into $z$. Conceptually, the latent should control sample diversity and the label should encode measurement constraints; entanglement between them undermines compliance and motivates enforcing their statistical independence during training.

To mitigate this root cause, we encourage statistical independence between the latent and label by matching the joint distribution $P(z,y)$ to the product $P(z)P(y)$. While these distributions are intractable to compute exactly, they are easy to sample: forming the paired set
\begin{equation}\label{eq:04Sset}
\SSS = \big\{(z_i,\, y_i)\big\}_{i=1}^{N}
\end{equation}
captures draws from $P(z,y)$ by encoding each sample $x_i$ to obtain $(z_i,y_i)$. Shuffling labels produces an empirical product set
\begin{equation}\label{eq:04Stildeset}
	\tilde{\SSS} = \big\{(z_i,\, \tilde{y}_i)\big\}_{i=1}^{N}
\end{equation}
that approximates samples from $P(z)P(y)$ for large $N$, where $(\tilde{y}_1,\ldots,\tilde{y}_N)$ is a random permutation of $(y_1,\ldots,y_N)$. We then minimize a sliced Wasserstein discrepancy between these sets~\citep{kolouri2019gsw}:
\begin{equation}\label{eq:04swloss}
\mathcal{L}^{\text{comp}} := \SW(\SSS,\,\tilde{\SSS}).
\end{equation}
This compliance loss is parameterized through the encoder outputs $z$ and is added to the standard loss function~\ref{eq:04trnloss} with an appropriate weight during training, improving label compliance at generation time.

\section{Results}\label{sec:05results}

We first present end-to-end inference examples that, given a single
label, generate ensembles of aerosol states, compute diagnostics, and
evaluate label compliance (Section~\ref{sec:051examples}). We then
aggregate results over the test set to summarize compliance errors,
generated-versus-true diagnostic scatter, and generative ambiguity for
both low- and high-dimensional label setups
(Section~\ref{sec:052smry}; see also Table~\ref{tab:01genambig}).

\subsection{Aerosol Diagnostic Inference Examples}\label{sec:051examples}


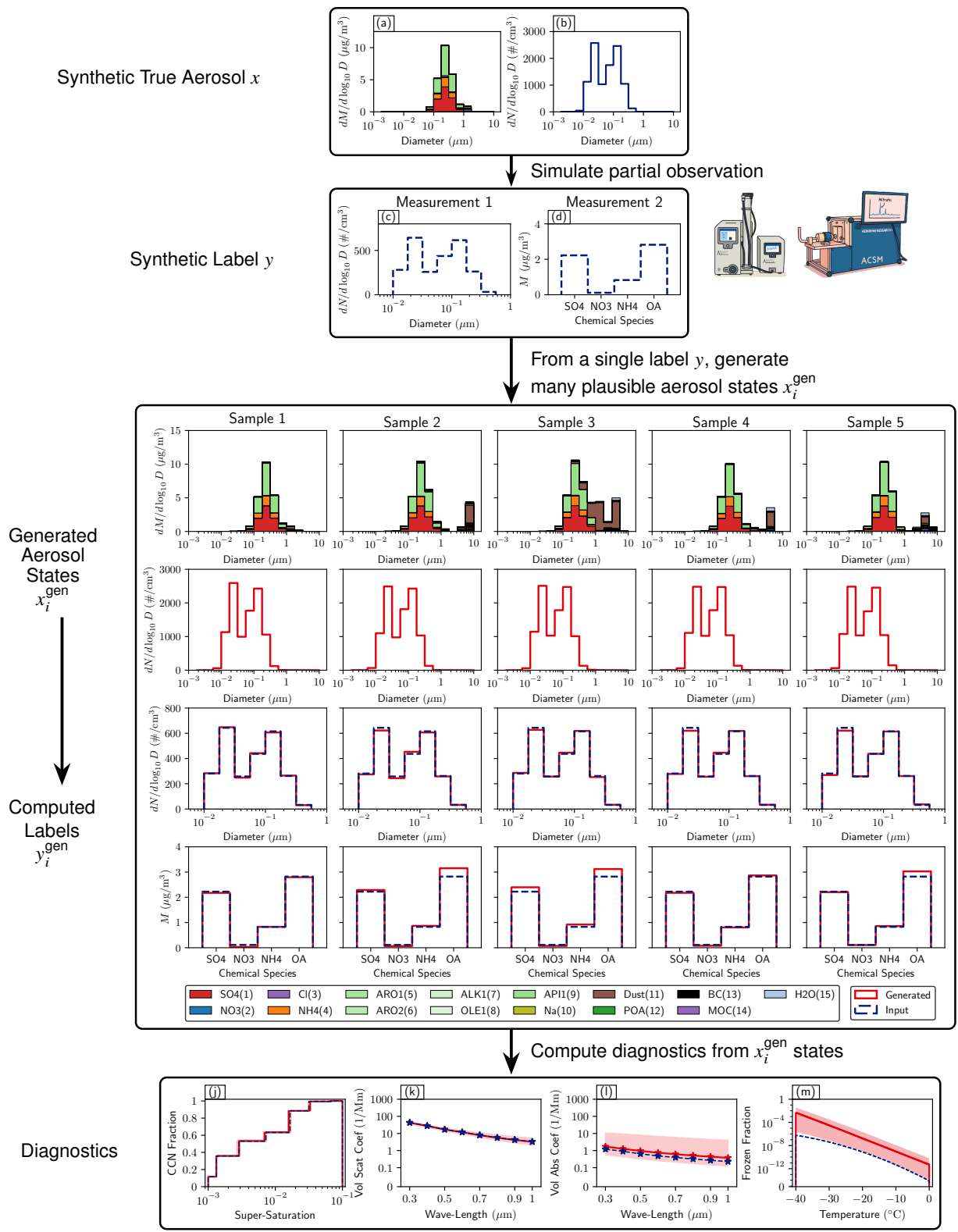
\begin{figure}
	\centering
	\scalebox{.95}{
		\begin{overpic}[height=0.97\textheight]{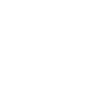} 
		\put(26,88){
			\begin{tcolorbox}[arc=1mm, 
				colback=white!0, 
				colframe=black!100, 
				width=0.39\linewidth, 
				left=1mm, right=1mm, 
				top=1mm, bottom=1mm,
				boxsep=0pt,
				boxrule=1pt]
				\includegraphics[page=1,width=1\textwidth]{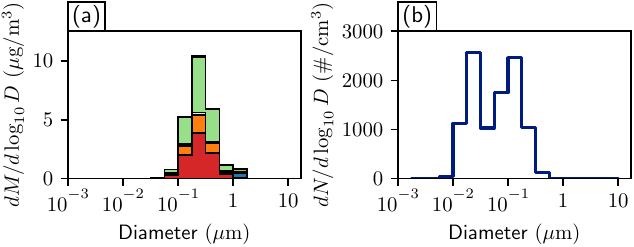}
			\end{tcolorbox}
		}
		\put(4,94){\sffamily\normalsize Synthetic True Aerosol $x$}
		
		\put(41,88){
			\begin{tikzpicture}[overlay, x=1mm, y=1mm]
			\draw[-{Stealth[length=3mm, width=2mm]}, line width=1.5pt, black]
			(0,0) -- (0, -5.5);
			\end{tikzpicture}
		}
		\put(43,86.25){\sffamily\normalsize Simulate partial observation}
		\put(58,78){\includegraphics[width=0.08\linewidth]{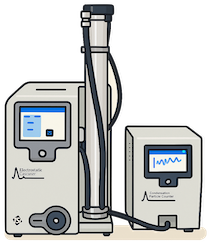}}
		\put(65,78){\includegraphics[width=0.12\linewidth]{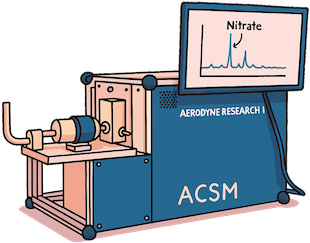}}

		\put(26,73){
			\begin{tcolorbox}[arc=1mm, 
				colback=white!0, 
				colframe=black!100, 
				width=0.39\linewidth, 
				left=1mm, right=1mm, 
				top=1mm, bottom=1mm,
				boxsep=0pt,
				boxrule=1pt]
				\includegraphics[page=1,width=1\textwidth]{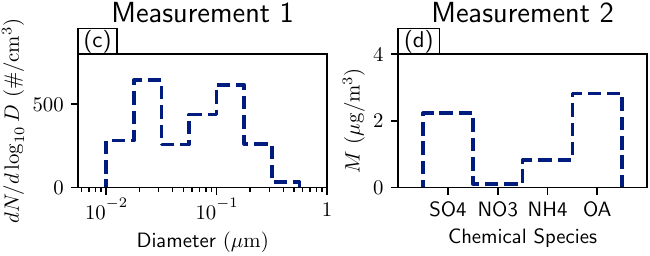}
			\end{tcolorbox}
		}
		\put(10,79){\sffamily\normalsize Synthetic Label $y$}
		
		\put(41,73){
			\begin{tikzpicture}[overlay, x=1mm, y=1mm]
			\draw[-{Stealth[length=4mm, width=3mm]}, line width=1.5pt, black] (0,0) -- (0, -11.75);
			\end{tikzpicture}
		}
		\put(43,68.5){\sffamily\normalsize\shortstack[l]{From a single label $y$, generate\\many plausible aerosol states $\gen{x}_i$}}

		\put(10,16){
			\begin{tcolorbox}[arc=1mm, 
				colback=white!0, 
				colframe=black!100, 
				width=0.89\linewidth, 
				left=1mm, right=1mm, 
				top=1mm, bottom=1mm,
				boxsep=0pt,
				boxrule=1pt]
				\includegraphics[page=1,width=1\textwidth]{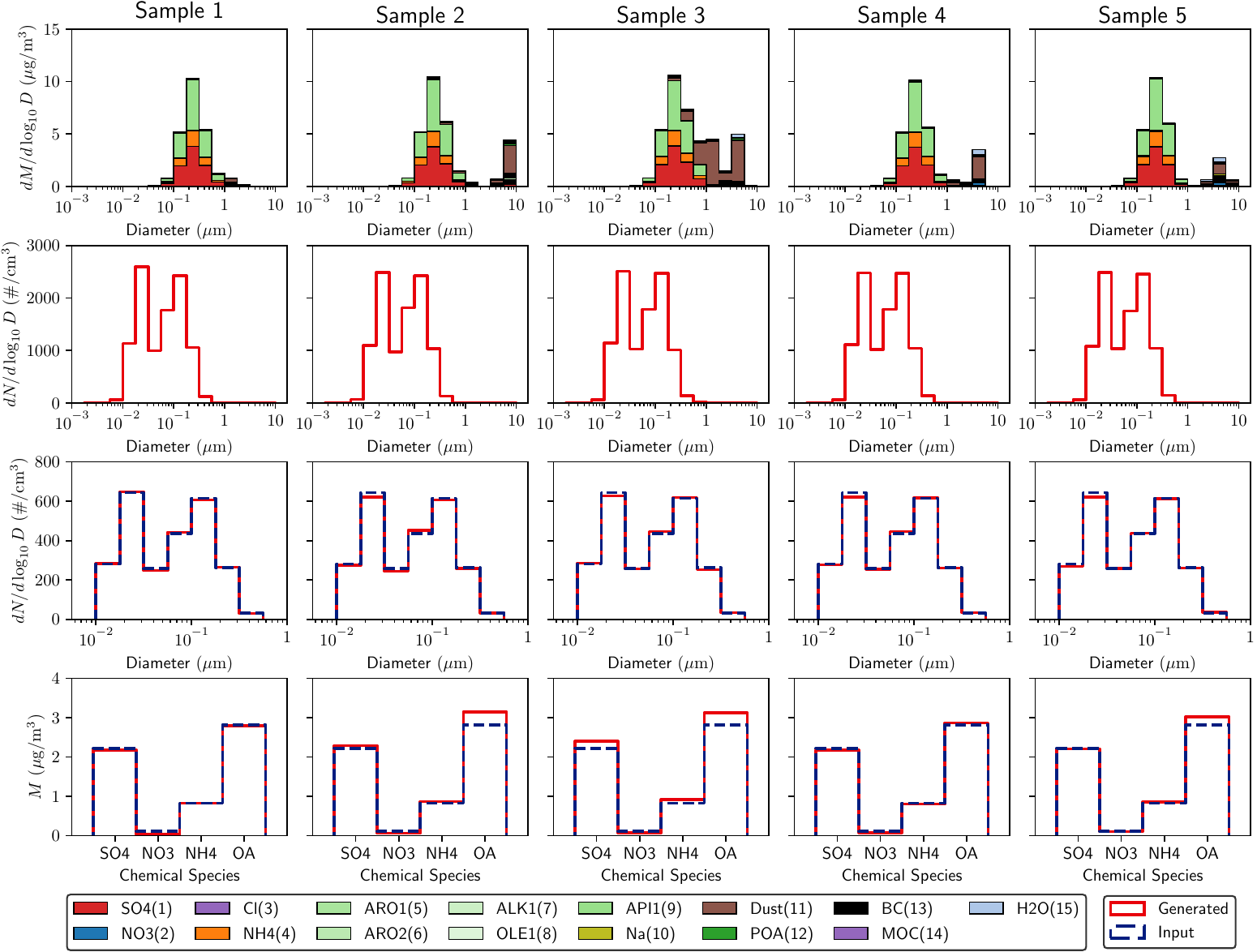}
			\end{tcolorbox}
		}
		\put(0,51){\sffamily\normalsize \shortstack[c]{Generated\\Aerosol \\States\\$\gen{x}_i$}}
		\put(4,50){
			\begin{tikzpicture}[overlay, x=1mm, y=1mm]
			\draw[-{Stealth[length=4mm, width=3mm]}, line width=1.5pt, black]
			(0,0) -- (0, -30);
			\end{tikzpicture}
		}
		\put(0,30){\sffamily\normalsize \shortstack[c]{Computed\\Labels\\$\gen{y}_i$}}

		\put(41,16){
			\begin{tikzpicture}[overlay, x=1mm, y=1mm]
			\draw[-{Stealth[length=4mm, width=3mm]}, line width=1.5pt, black] 
			(0,0) -- (0, -8);
			\end{tikzpicture}
		}
		\put(43,13.75){\sffamily\normalsize Compute diagnostics from $\gen{x}_i$ states}

		\put(12,-0.5){
			\begin{tcolorbox}[arc=1mm, 
				colback=white!0, 
				colframe=black!100, 
				width=0.8475\linewidth, 
				left=1mm, right=1mm, 
				top=1mm, bottom=1mm,
				boxsep=0pt,
				boxrule=1pt]
				\includegraphics[page=1,width=1\textwidth]{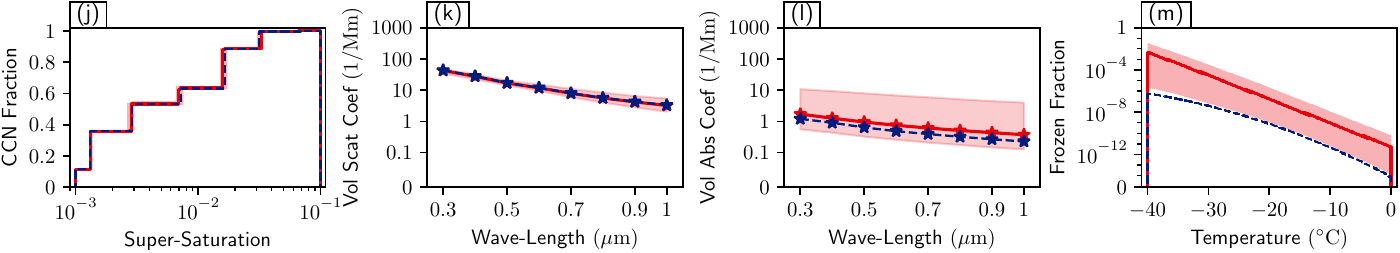}
			\end{tcolorbox}
			}
		\put(1,5.5){\sffamily\normalsize Diagnostics}
		\end{overpic}
	}
	\vspace{-4mm}\caption{\textbf{Low-dimensional label setup:} An example of estimating the aerosol state.}
	\label{fig:03ldmflowchart}
\end{figure}


\begin{figure}
	\centering
	\scalebox{.95}{
		\begin{overpic}[height=0.97\textheight]{figures/05_blank.png} 
		\put(26,88){
			\begin{tcolorbox}[arc=1mm, 
				colback=white!0, 
				colframe=black!100, 
				width=0.39\linewidth, 
				left=1mm, right=1mm, 
				top=1mm, bottom=1mm,
				boxsep=0pt,
				boxrule=1pt]
				\includegraphics[page=1,width=1\textwidth]{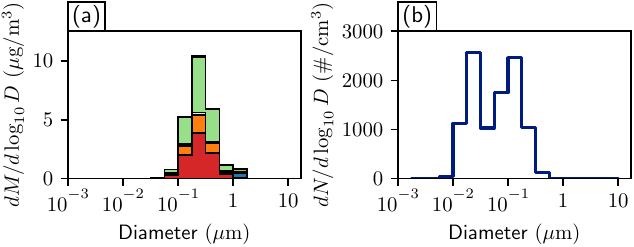}
			\end{tcolorbox}
		}
		\put(4,94){\sffamily\normalsize Synthetic True Aerosol $x$}
		
		\put(41,88){
			\begin{tikzpicture}[overlay, x=1mm, y=1mm]
			\draw[-{Stealth[length=3mm, width=2mm]}, line width=1.5pt, black]
			(0,0) -- (0, -5.5);
			\end{tikzpicture}
		}
		\put(43,86.25){\sffamily\normalsize Simulate partial observation}

		\put(19.5,73){
			\begin{tcolorbox}[arc=1mm, 
				colback=white!0, 
				colframe=black!100, 
				width=0.56\linewidth, 
				left=1mm, right=1mm, 
				top=1mm, bottom=1mm,
				boxsep=0pt,
				boxrule=1pt]
				\includegraphics[page=1,width=1\textwidth]{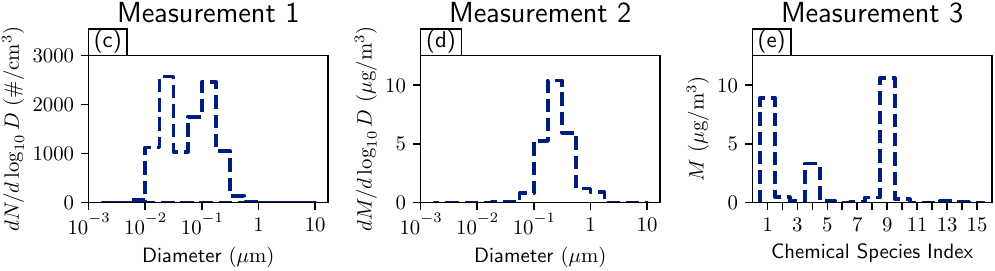}
			\end{tcolorbox}
		}
		\put(3.5,79){\sffamily\normalsize \shortstack[l]{Synthetic Label $y$}}
		
		\put(41,73){
			\begin{tikzpicture}[overlay, x=1mm, y=1mm]
			\draw[-{Stealth[length=4mm, width=3mm]}, line width=1.5pt, black]
			(0,0) -- (0, -11.75);
			\end{tikzpicture}
		}
		\put(43,68.5){\sffamily\normalsize\shortstack[l]{From a single label $y$, generate\\many plausible aerosol states $\gen{x}_i$}}

		\put(10,16){
			\begin{tcolorbox}[arc=1mm, 
				colback=white!0, 
				colframe=black!100, 
				width=0.8975\linewidth, 
				left=1mm, right=1mm, 
				top=1mm, bottom=1mm,
				boxsep=0pt,
				boxrule=1pt]
				\includegraphics[page=1,width=1\textwidth]{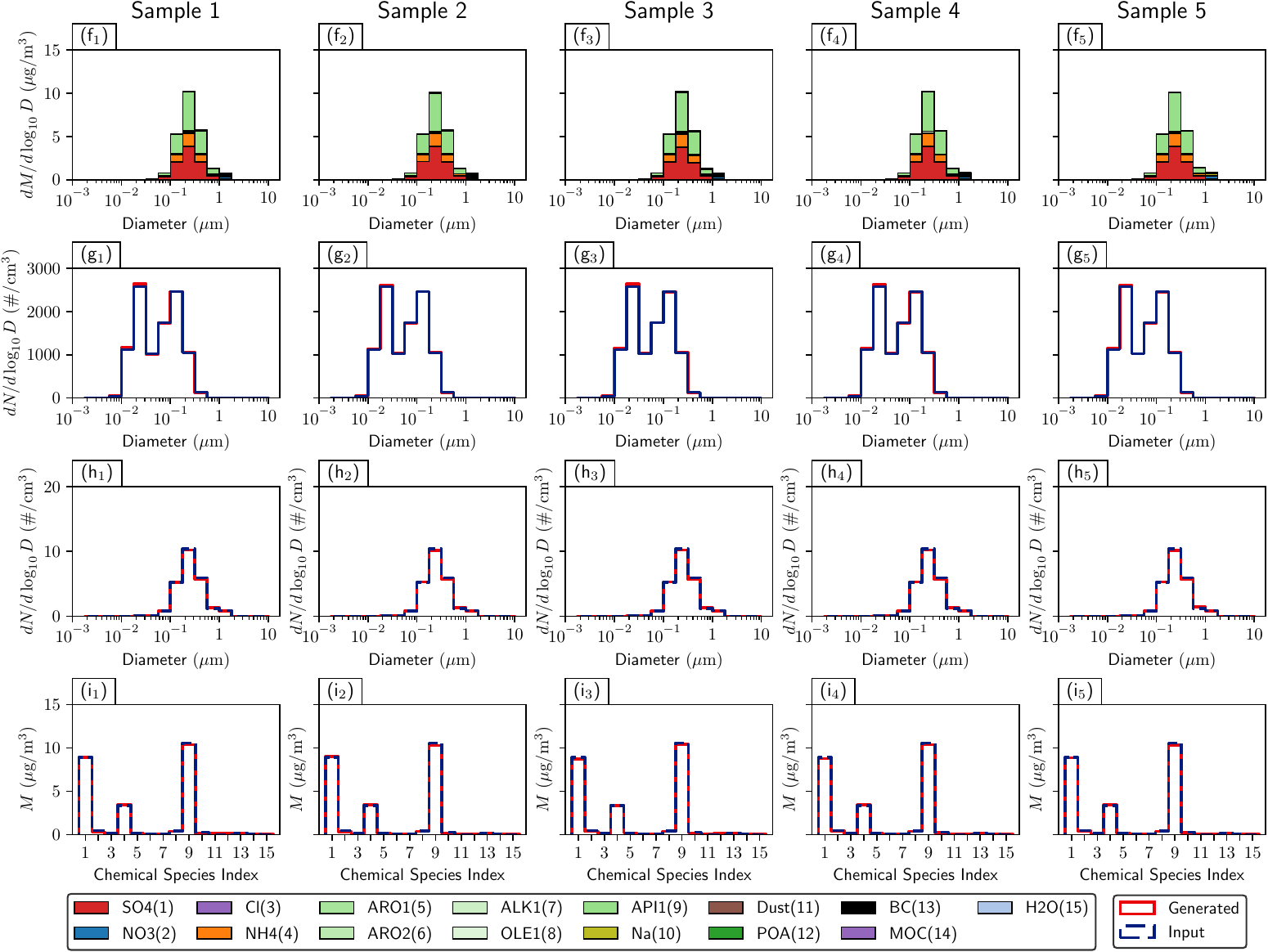}
			\end{tcolorbox}
		}

		\put(0,51){\sffamily\normalsize \shortstack[c]{Generated\\Aerosol\\States\\$\gen{x}_i$}}
		\put(4,50){
			\begin{tikzpicture}[overlay, x=1mm, y=1mm]
			\draw[-{Stealth[length=4mm, width=3mm]}, line width=1.5pt, black]
			(0,0) -- (0, -30);
			\end{tikzpicture}
		}
		\put(0,30){\sffamily\normalsize \shortstack[c]{Computed\\Labels\\$\gen{y}_i$}}

		\put(41,16){
			\begin{tikzpicture}[overlay, x=1mm, y=1mm]
			\draw[-{Stealth[length=4mm, width=3mm]}, line width=1.5pt, black]
			(0,0) -- (0, -8);
			\end{tikzpicture}
		}
		\put(43,13.75){\sffamily\normalsize Compute diagnostics from $\gen{x}_i$ states}

		\put(12,-0.5){
			\begin{tcolorbox}[arc=1mm, 
				colback=white!0, 
				colframe=black!100, 
				width=0.8475\linewidth, 
				left=1mm, right=1mm, 
				top=1mm, bottom=1mm,
				boxsep=0pt,
				boxrule=1pt]
				\includegraphics[page=1,width=1\textwidth]{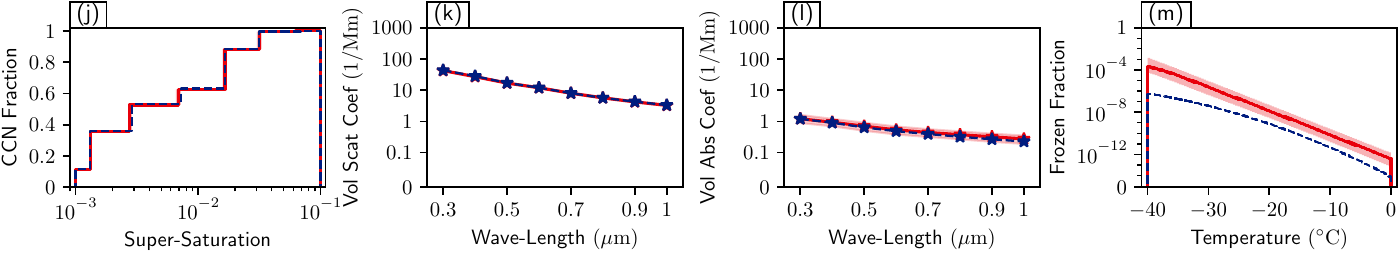}
			\end{tcolorbox}
			}
		\put(1,5.5){\sffamily\normalsize Diagnostics}
		\end{overpic}
	}
	\vspace{-4mm}\caption{\textbf{High-dimensional label setup:} An example of estimating the aerosol state.}
	\label{fig:04hdmflowchart}
\end{figure}

Figures~\ref{fig:03ldmflowchart} and~\ref{fig:04hdmflowchart} outline
the analysis pipeline: both begin with the same synthetic true aerosol
$x$ at the top, after which we compute labels as defined in
Section~\ref{sec:033aeromeasures}. Figure~\ref{fig:03ldmflowchart}
shows the low-dimensional label setup with 11
dimensions, while Figure~\ref{fig:04hdmflowchart} employs the
high-dimensional label setup with 55 dimensions.

For each input label $y$, we generate an ensemble of 100 conditional
aerosol states and visualize five examples to illustrate
variability. In the low-dimensional label setup (Figure~\ref{fig:03ldmflowchart}),
coarse-mode dust varies widely because the label does not include
information about refractory dust.  By contrast, the high-dimensional
label setup (Figure~\ref{fig:04hdmflowchart}) encodes stronger
constraints and yields markedly reduced inter-sample variation
(Section~\ref{sec:033aeromeasures}).

Below each set of generated aerosol samples, we plot the computed
label $\gen{y}_i$ in solid red and overlay the input label in dashed
blue to directly assess conditional fidelity.  In the low-dimensional label setup
(Figure~\ref{fig:03ldmflowchart}g), the number distribution components
agree nearly perfectly with the inputs, while bulk-composition
components show slightly larger yet still excellent agreement.  For
the high-dimensional label setup (Figure~\ref{fig:04hdmflowchart}), agreement is
near-perfect across all components, consistent with the stronger
constraints encoded by the higher-dimensional setup
(Section~\ref{sec:033aeromeasures}).

From each ensemble of 100 conditional samples $\gen{x}_i$, we compute
the full set of diagnostics (CCN, optical, and frozen fraction
spectra) and summarize them by the ensemble mean (solid red) with the
ensemble range shown as a red-shaded confidence interval in
Figures~\ref{fig:03ldmflowchart} and~\ref{fig:04hdmflowchart},
reflecting uncertainty from partial measurements and latent
variability in the conditional generator.

In the diagnostic plots, dashed blue curves denote the synthetic true
spectra, whereas solid red curves show the ensemble mean from
generated samples; exact overlap is not expected because partial
labels do not uniquely determine the full aerosol
state.  Instead, the blue curve should typically lie within the
red-shaded confidence interval, which represents the range of spectra
plausible under the given label.  For the
low-dimensional label setup, confidence intervals for frozen
fraction and absorption are wide because frozen fraction is sensitive
to dust and absorption to black carbon—quantities that the low-dimensional
labels do not tightly constrain.  By contrast, CCN and optical scattering
exhibit relatively narrow intervals, reflecting stronger constraints
from number-distribution and bulk-composition information in the
labels and accurate reconstruction of these diagnostics.

Comparing the low- and high-dimensional label setups, the shaded
confidence intervals are systematically narrower in the
high-dimensional setup, reflecting tighter constraints on dust and
black carbon (Section~\ref{sec:033aeromeasures}). In the
low-dimensional setup, by contrast, the labels do not constrain the
coarse mode, and the generative model sometimes produces coarse dust
particles. This outcome is not a hallucination but rather an
expression of epistemic uncertainty: in the absence of observational
information, the VAE draws on prior knowledge from training, where
many aerosol states included realistic coarse particles. The
appearance of such particles should therefore be interpreted as
plausible completions of the aerosol state given the limited constraints imposed by the labels.

This behavior has diagnostic-specific implications. For CCN activity,
which is dominated by fine particles, the additional coarse mode has
little influence. However, for diagnostics sensitive to dust, such as
immersion freezing, the “invented” coarse particles manifest as a
wider spread across generated states. In particular, the
frozen-fraction diagnostic remains more error-prone, both because the
training did not prioritize dust reconstruction and because this
diagnostic is highly sensitive to even small dust loads. This
limitation was also noted in~\citet{saleh2025generative} and is
consistent with the equal weighting of species in the reconstruction
loss. As discussed there, increasing the relative weight on dust
during training could improve frozen-fraction performance by
emphasizing the most influential species for this diagnostic.
	
\subsection{Collective Aerosol Diagnostic Summaries}\label{sec:052smry}


\begin{figure*}
	\centering
	\includegraphics[width=0.98\linewidth]{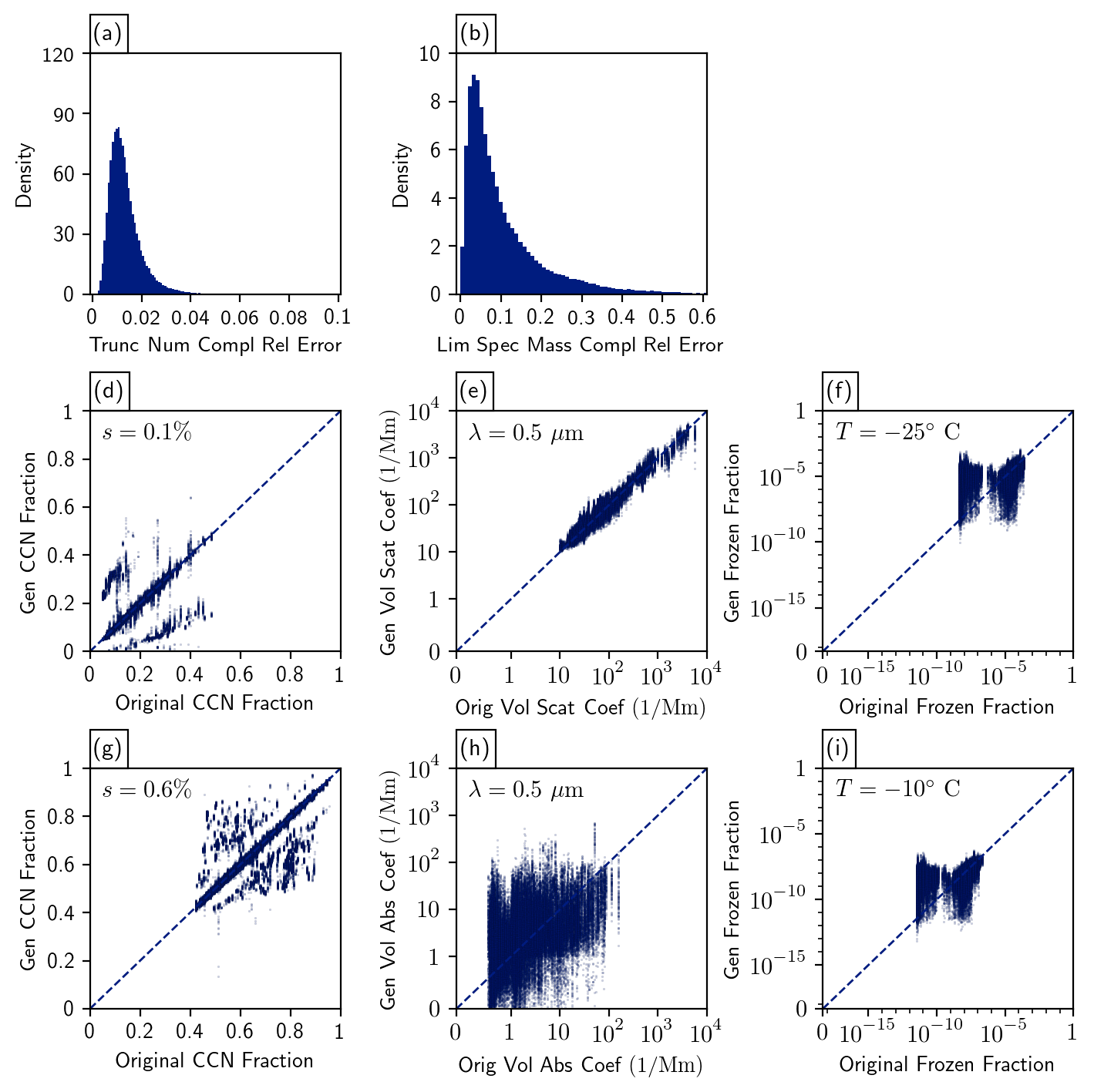}
	\caption{\textbf{Low-dimensional label setup:} The collective aerosol diagnostic summary plots on the testing portion of the data. 
	(a) and (b) show the truncated number distribution and limited species bulk mass label compliance errors, respectively.
	(d) and (g) show the input vs. generated CCN fraction scatter plots at $s=0.1\%$ and $s=0.3\%$ supersaturation levels, respectively. 
	(e) and (h) show the input vs. generated volume scattering and absorption coefficient scatter plots at $\lambda=0.5 \, {\rm \mu m}$ wavelength.
	(f) and (i) show the input vs. generated frozen fraction plots at $T=-25 \, {\rm ^{\circ} C}$ and $T=-10 \, {\rm ^{\circ} C}$, respectively.}
	\label{fig:05ldmsmry}
\end{figure*}


\begin{figure*}
	\centering
	\includegraphics[width=0.98\linewidth]{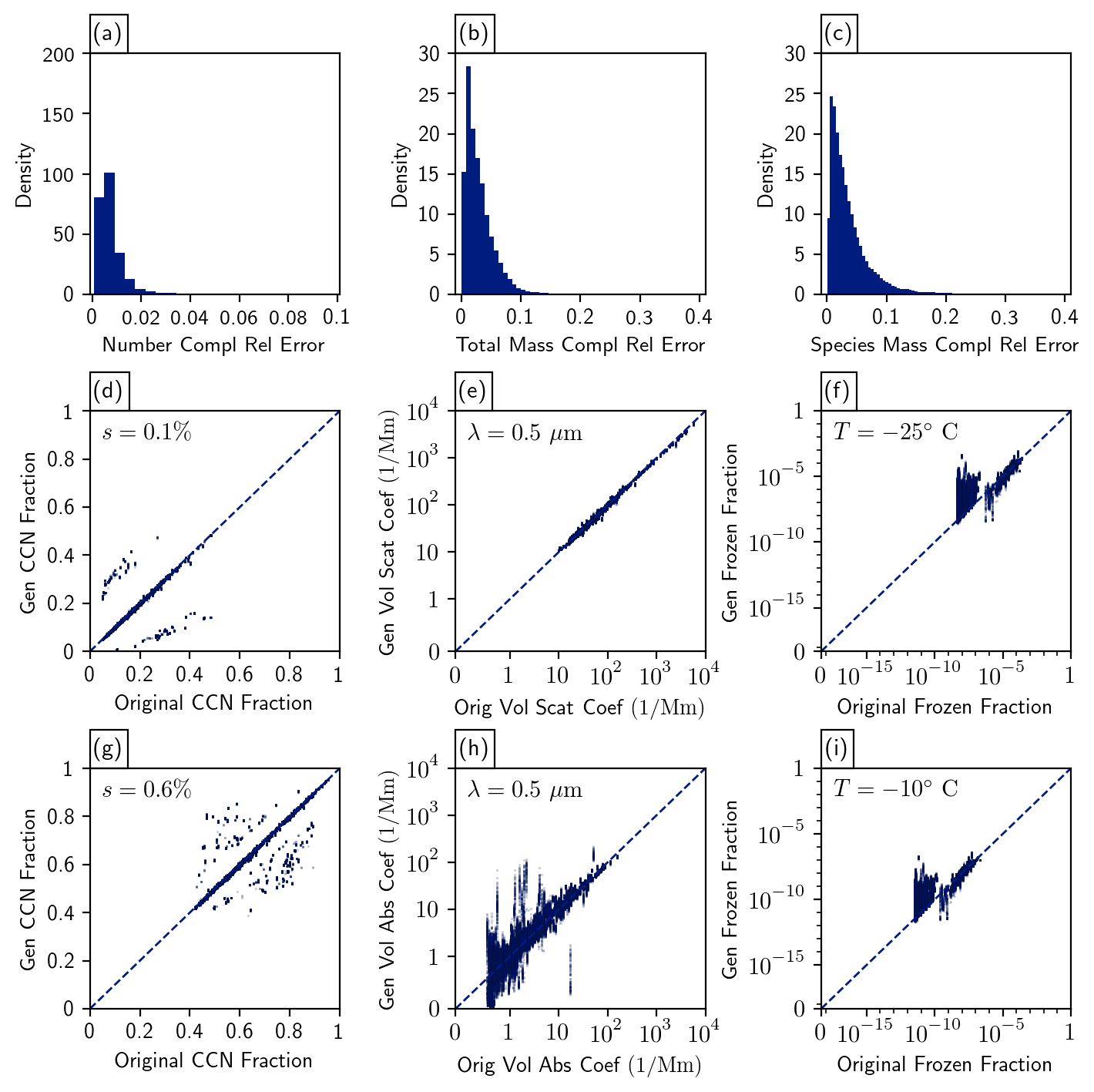}
	\caption{\textbf{High-dimensional label setup:} The collective aerosol diagnostic summary plots on the testing portion of the data. 
	(a), (b), and (c) show the number, total mass and species mass label compliance errors, respectively.
	(d) and (g) show the input vs. generated CCN fraction scatter plots at $s=0.1\%$ and $s=0.3\%$ supersaturation levels, respectively. 
	(e) and (h) show the input vs. generated volume scattering and absorption coefficient scatter plots at $\lambda=0.5 \, {\rm \mu m}$ wavelength.
	(f) and (i) show the input vs. generated frozen fraction plots at $T=-25 \, {\rm ^{\circ} C}$ and $T=-10 \, {\rm ^{\circ} C}$, respectively.}
	\label{fig:06hdmsmry}
\end{figure*}

Figures~\ref{fig:05ldmsmry} and~\ref{fig:06hdmsmry} present collective summary
plots for the low- and high-dimensional label setups. The top panels show
histograms of label compliance errors, defined as the relative error between the
input $y$ and the generated $\gen{y}_i$ computed from conditional samples
$\gen{x}_i$. This quantity corresponds to the pointwise difference between the
solid red (computed) and dashed blue (input) curves in the computed measurement
panels of Figures~\ref{fig:03ldmflowchart} and~\ref{fig:04hdmflowchart}, serving
as a post-generation consistency check that decoded samples preserve the
specified labels. Ideally these errors are zero and, for a well-trained
conditional model, should be small. In the low-dimensional label setup, the
average compliance errors are 1.3\% for the truncated number distribution and
10.8\% for the limited species bulk masses, while in the high-dimensional setup
the number-distribution, total-mass, and species-mass errors are 0.78\%, 3.07\%,
and 3.79\%, respectively. These low values indicate accurate label
reconstruction and confirm that the generated samples remain relevant to the
conditioning information.

The scatter plots compare diagnostics from generated samples $\gen{x}_i$
($y$-axis) against those from the simulated true aerosol $x$ ($x$-axis); points
correspond to values taken from the red-shaded ranges plotted against the blue
dashed curve in panels (j--m) of Figures~\ref{fig:03ldmflowchart}
and~\ref{fig:04hdmflowchart}. Because the labels only partially constrain the
underlying state, points need not fall on the one-to-one line---spread reflects
the inherent non-uniqueness of the inference problem. Nevertheless, the true
values should lie within the vertical spread for each $x$ value, since the true
state is a valid member of the conditional distribution, meaning the one-to-one
line should lie within the range of points.

Comparing the summary plots, CCN and scattering (Figure~\ref{fig:05ldmsmry}j,k)
are most tightly constrained, whereas frozen fraction and absorption remain less
well constrained. Moreover, the high-dimensional-label scatter plots exhibit
reduced variability relative to the low-dimensional setup, with points more
tightly clustered around the one-to-one line in Figure~\ref{fig:06hdmsmry}(d,g)
than in Figure~\ref{fig:05ldmsmry}(d,g), reflecting stronger measurement
constraints in the high-dimensional setup.

\renewcommand{\arraystretch}{1}
\newcommand{\ZR}{\phantom{0}}
\begin{table}[t]
	\smaller[0.5] 
	\centering	
	\begin{tabular}{p{0.30\linewidth} >{\centering\arraybackslash}p{0.3\linewidth} >{\centering\arraybackslash}p{0.3\linewidth}}
	\hline
	Generative Ambiguity & \makecell[c]{Low-Dim Measurements\\Mean [95\% CI]} & \makecell[c]{High-Dim Measurements\\Mean [95\% CI]} \\\hline
	Truncated Number Distribution & \ZR1.33\% [\ZR1.29\%, \ZR1.37\%] & \textbf{\ZR0.78\%} [\ZR0.72\%, \ZR0.87\%] \\
	Limited Species Bulk Masses & 10.8\ZR\% [10.2\ZR\%, 11.3\ZR\%] & \textbf{\ZR5.40\%} [\ZR4.82\%, \ZR6.18\%] \\
	Number Distribution & \ZR1.49\% [\ZR1.44\%, \ZR1.55\%] & \textbf{\ZR0.78\%} [\ZR0.72\%, \ZR0.87\%] \\
	Total Mass Distribution & 33.9\ZR\% [33.0\ZR\%, 35.0\ZR\%] & \textbf{\ZR3.07\%} [\ZR2.84\%, \ZR3.33\%] \\
	Species Bulk Masses & 34.9\ZR\% [33.8\ZR\%, 35.8\ZR\%] & \textbf{\ZR3.79\%} [\ZR3.66\%, \ZR4.03\%] \\
	CCN Spectrum & \ZR5.20\% [\ZR5.11\%, \ZR5.32\%] & \textbf{\ZR3.34\%} [\ZR3.18\%, \ZR3.48\%] \\
	Scattering Log-Spectrum & \ZR2.69\% [\ZR2.59\%, \ZR2.81\%] & \textbf{\ZR0.86\%} [\ZR0.83\%, \ZR0.93\%] \\
	Absorption Log-Spectrum & 44.2\ZR\% [43.5\ZR\%, 45.0\ZR\%] & \textbf{17.0\ZR\%} [15.9\ZR\%, 18.2\ZR\%] \\
	Frozen Fraction Log-Spectrum & \ZR7.87\% [\ZR7.55\%, \ZR8.14\%] & \textbf{\ZR3.59\%} [\ZR3.41\%, \ZR3.81\%] \\\hline
	\end{tabular}
	\caption{The generative ambiguity of the low vs. high-dimensional measurement settings. 
	As expected, the generative ambiguity of LDM settings is significantly higher than those of the HDM settings.}
	\label{tab:01genambig}
\end{table}

Table~\ref{tab:01genambig} summarizes generative ambiguity across measurements
and diagnostics, defined for each diagnostic as the mean relative error between
values from the synthetic true aerosol and those from generated samples
conditioned only on the labels. That is, given a diagnostic $\gamma$ computed
from the true state $x$, and $\gen{\gamma}_i$ being the corresponding diagnostic
computed from the generated sample $\gen{x}_i$, we define the generative ambiguity
for this true state as
\begin{equation}\label{eq:06genambig}
	\text{Generative Ambiguity} = \frac{1}{N} \sum_{i=1}^{N} \frac{\|\gamma - \gen{\gamma}_i\|_2}{\|\gamma\|_2 + \|\gen{\gamma}_i\|_2}.
\end{equation}
In Table~\ref{tab:01genambig} we list the generative ambiguity averaged over all
true states $x$.

For measured quantities, generative ambiguity~(\ref{eq:06genambig}) coincides
with the compliance error~(\ref{eq:05comperr}) and should approach zero for a
well-trained conditional model; for diagnostics not directly constrained,
nonzero ambiguity is expected because the labels do not uniquely determine the
diagnostic. Lower values indicate stronger conditional fidelity, and, as
expected, the high-dimensional label setup exhibits uniformly lower ambiguity
than the low-dimensional setup.

Some visual artifacts in the scatter summaries of
Figures~\ref{fig:05ldmsmry} and~\ref{fig:06hdmsmry} reflect
reconstruction precision limits of the underlying autoencoder. CCN
fractions and volume scattering are recovered most readily, with minor
shifts arising from the binned state representation (see~\citet{saleh2025generative}). Scattering is easier to reproduce than absorption because
many species contribute to scattering, whereas absorption is dominated
by BC. Frozen fraction remains the most challenging diagnostic with
occasional overestimation of dust-free cases (see~\citet{saleh2025generative}).

Overall, these experiments demonstrate that while low-dimensional
labels are sufficient to recover CCN and scattering with good
fidelity, high-dimensional labels markedly improve constraints on
absorption and frozen fraction diagnostics.

\section{Conclusions}

We introduced a conditional generative framework that maps a label to an
ensemble of plausible aerosol states and propagates this ensemble to diagnostics
(CCN, optical scattering/absorption, frozen fraction) to provide mean estimates
with confidence intervals. A CVAE backbone with a Wasserstein-based
regularization encouraging label-latent independence yields small compliance
errors, ensuring generated samples remain consistent with the conditioning
information.

Across the different label setups, high-dimensional labels (including full
number and total mass distributions plus all species bulk masses) substantially
reduce variability and generative ambiguity relative to low-dimensional labels,
while the low-dimensional setting still constrains CCN and scattering well but
underconstrains dust- and BC-sensitive diagnostics. These findings clarify which
observational inputs most effectively constrain different climate-relevant
properties and illustrate the value of systematic comparisons between
alternative measurement configurations.

Beyond the proof-of-concept presented here, the framework admits several modes
of application depending on the availability of measurements and simulations.
(1) Fully measurement-based training and inference: both states and labels are
drawn from observations. This would constitute a truly model-free approach,
capable of learning directly from data, though it requires extensive and diverse
training datasets.  (2) Hybrid training and inference: simulated states and
labels are used for training, optionally with fine-tuning on observational data,
while real measurements provide labels for inference. This configuration is
practical in the near term, though care must be taken to address potential
distribution shifts between simulated and observed populations. (3) Fully
simulated training and inference: both training and inference rely on simulated
states and labels. This is the setting we adopt here to demonstrate the method
under controlled conditions.  Together, these variants highlight the flexibility
of the generative framework and point toward future opportunities for
integrating simulated and real-world data to maximize the value of aerosol
observations.

More broadly, this work demonstrates a pathway for translating partial
observations into actionable diagnostic estimates with quantified uncertainty,
without the need for custom forward-inverse models. The approach is modular with
respect to measurement definitions and diagnostics, making it extensible to a
wide range of observational setups and targets. While our study relies entirely
on synthetic data, the framework is designed for eventual integration with
real-world measurements, where it could inform instrument deployment strategies,
optimize field campaign design, and help close persistent gaps in
aerosol-climate constraints. Beyond aerosols, the methodology illustrates how
generative machine learning can complement process-based modeling across the
Earth sciences, offering new ways to leverage incomplete observations for robust
scientific inference.

\clearpage
\acknowledgments
This work used GPU resources at the Delta supercomputer of the National Center for Supercomputing Applications through Allocation CIS220111 from the Advanced Cyberinfrastructure Coordination Ecosystem: Services and Support (ACCESS) program~\citep{boerner2023access}, which is supported by National Science Foundation grants \#2138259, \#2138286, \#2138307, \#2137603, and \#2138296.

This work was supported by the U.S. Department of Energy, Office of Science, Office of Biological and Environmental Research under Award Number DE-SC0022130, and the Laboratory Directed Research and Development program at Sandia National Laboratories. Sandia National Laboratories is a multimission laboratory managed and operated by National Technology and Engineering Solutions of Sandia LLC, a wholly owned subsidiary of Honeywell International Inc. for the U.S. Department of Energy’s National Nuclear Security Administration contract DE-NA0003525.

This paper describes objective technical results and analysis. Any subjective views or opinions that might be expressed in the paper do not necessarily represent the views of the U.S. Department of Energy or the United States Government.

%
%
\datastatement
The underlying data for this study can be accessed at \url{https://doi.org/10.13012/B2IDB-2774261_V1}. The code used for the analysis is available at \url{https://github.com/ehsansaleh/partnn}.

\newpage
\appendix

\subsection{Probabilistic and Mathematical Notations} 
\newcommand{\zz}{z}
We denote expectations with $\EE_{P(\zz)}[h(\zz)]:=\int_{\zz} h(\zz)P(\zz)\diff \zz$. Note that only the random variable in the subscript (i.e., $\zz$) is eliminated after the expectation. The set of samples $\{u_1, \cdots u_N\}$ is denoted with $\{u_i\}_{i=1}^N$. We write $h_{\theta}(u)$ to denote the output of a neural network, parameterized by $\theta$, on the input $u$. These notations and operators are summarized in Table~\ref{tab:mathnotation}.

\renewcommand{\arraystretch}{1.2}
\begin{table}[h]
	\centering	
	\begin{tabular}{p{0.15\textwidth}p{0.75\textwidth}}
	    \hline
		Notation & Description \\ \hline
		$f_{\theta}$ & Encoder network parameterized by $\theta$ \\
		$g_{\theta}$ & Decoder network parameterized by $\theta$ \\
		$N$ & Number of samples \\
		$m$ & Speciated mass distribution \\
		$n$ & Number distribution \\
		$x$ & True aerosol state consisting of $m$ and $n$\\
		$y$ & Label for a true aerosol state $x$\\
		$\TTx$ & Aerosol state preprocessing transformation \\
		$\TTy$ & Label preprocessing transformation \\
		$\mu$ & Variational latent mean variable for a sample \\
		$\Sigma$ & Variational latent covariance variables for a sample \\
		$z$ & Latent variable representation \\
		$s$ & Critical relative humidity super-saturation level \\
		$\lambda$ & Optical properties wave-length \\
		$\{(u_i, v_i)\}_{i=1}^N$ & Generic paired sample set definition \\
		$\EE_{P(\zz)}[h(\zz)]$ & Expectation of $h(z)$ over $z\sim P(\cdot)$\\
		$\SW$ & Sliced Wasserstein distance \\
		$\gen{z}$ & Generative latent variable sampled from $\mathcal{N}(0,I)$\\
		$\gen{x}$ & Generated aerosol state \\
		$\gen{y}$ & Computed labels for the generated aerosol state\\
		\hline
	\end{tabular}	
	\caption{The mathematical notations used throughout the paper.}
	\label{tab:mathnotation}
\end{table}

\begin{figure*}[t]
	\centering
	\includegraphics[page=3,width=0.98\linewidth]{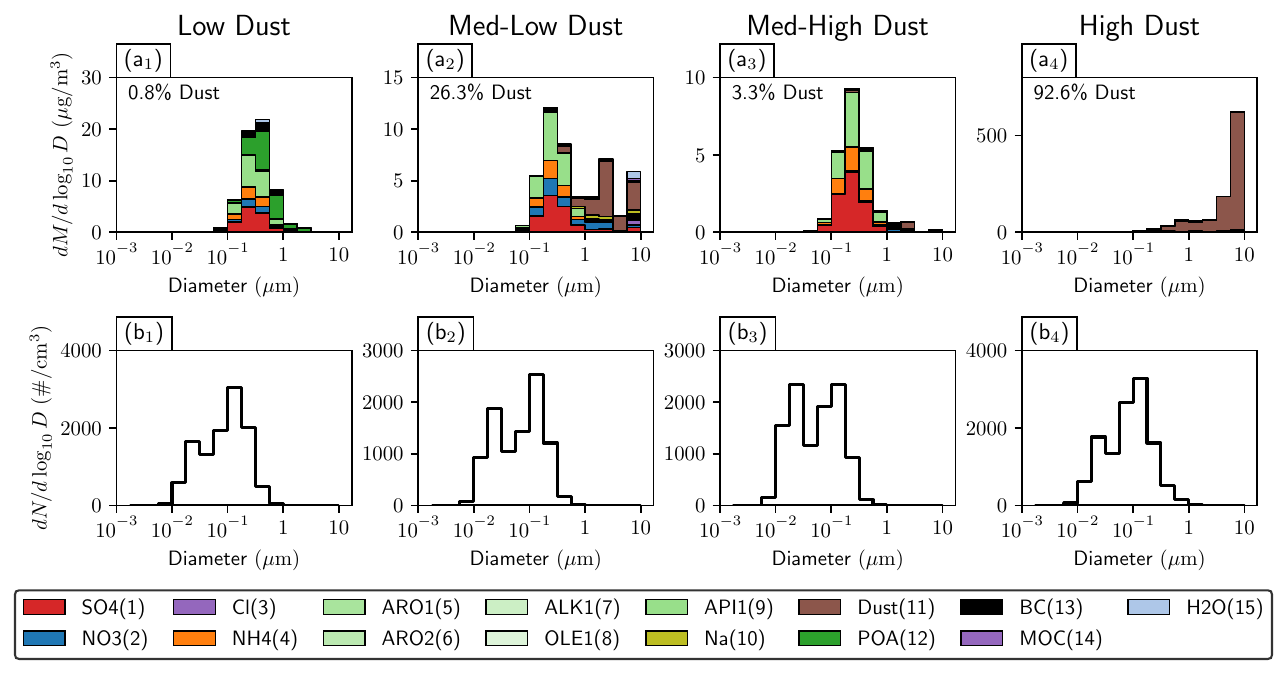}
	\caption{\textbf{Traditional CVAE:} Examples generated from the low to high dust fraction classes.}
	\label{fig:a01tradanec}
\end{figure*}

\begin{figure*}
	\centering
	\includegraphics[page=11,width=0.98\linewidth]{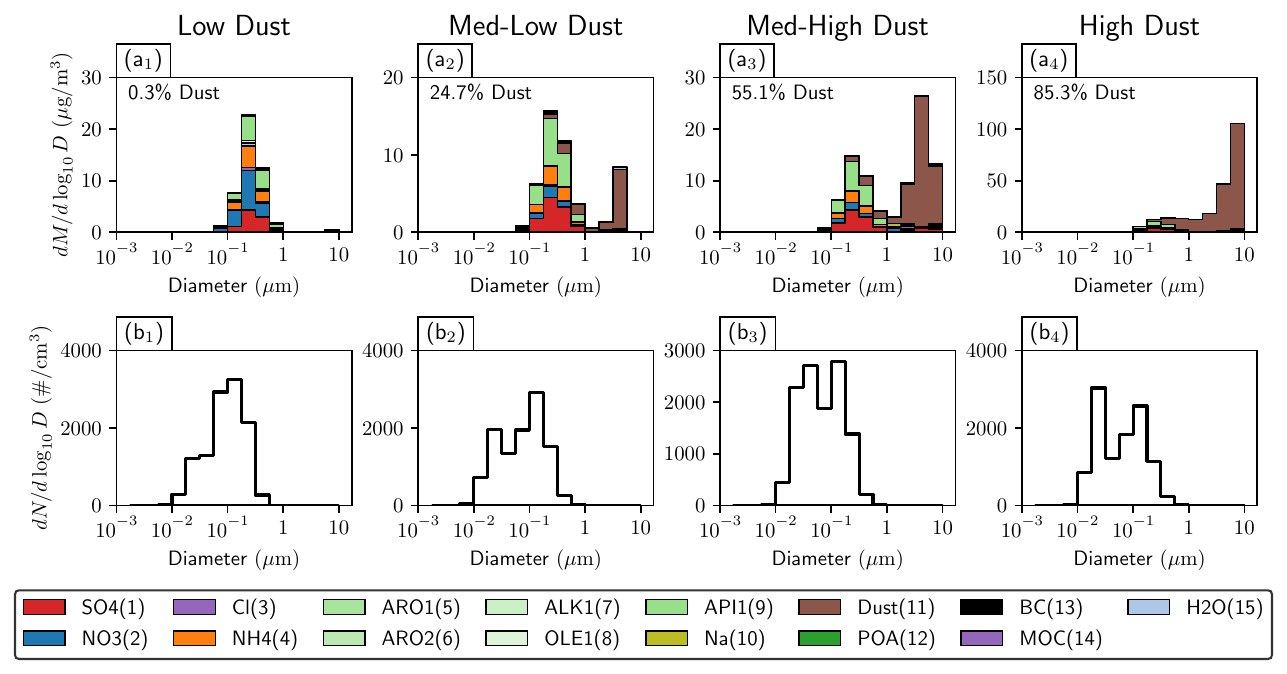}
	\caption{\textbf{Wasserstein-regularized CVAE:} Examples generated from the low to high dust fraction classes.}
	\label{fig:a02wrcvaeanec}
\end{figure*}

\subsection{Implementation Details}

We used the Adam~\citep{kingma2014adam} optimizer with a learning rate of $10^{-3}$. The models were trained for 10000 iterations with a mini-batch size of 2048. We used a fully-connected network with 2 layers, 128 hidden units, and the ReLU activation function. For the optical properties calculations, we used 220 iterations in the Toon-Ackerman algorithm.

\subsection{Impact of Wasserstein Regularization: Categorical Labels}\label{sec:a02catlbl}

To stress-test conditional generation under weak constraints, we consider a setting where the input label is categorical and the true states are highly diverse. This setting's labels partition samples by dust mass fraction into four non-overlapping categories: Low Dust (0--4\%; lower 40th percentile of the training data), Medium-Low Dust (4--55\%; 40th--60th percentiles of the training data), Medium-High Dust (55--84\%; 60th--80th percentiles of the training data), and High Dust (84--100\%; 80th--100th percentiles of the training data). Each training sample belongs to exactly one category. The modeling challenge is to generate only samples that belong to the specified input category.

Figures~\ref{fig:a01tradanec} and~\ref{fig:a02wrcvaeanec} compare anecdotal generations from a traditional CVAE and our Wasserstein-regularized CVAE. In Figure~\ref{fig:a01tradanec}, the medium-high dust example exhibits a smaller dust fraction than the medium-low dust example, indicating a failure to comply with the input label. By contrast, Figure~\ref{fig:a02wrcvaeanec} shows samples from our method in which this label-compliance issue is less pronounced.

\begin{figure*}
	\centering
	\begin{overpic}[width=1\linewidth]{figures/05_blank.png}
	\put(0,91){\includegraphics[width=0.98\linewidth,trim=0 175 0 0,clip]{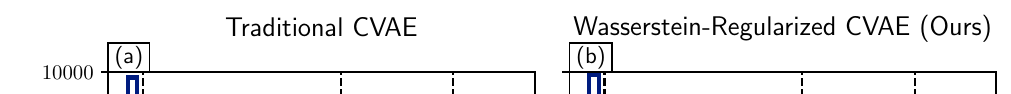}}
	\put(0,55.5){\includegraphics[width=0.98\linewidth,trim=0 0 0 52,clip]{figures/36_frachistcmprd.pdf}}
	\put(10.2,90.25){/}
	\put(51.5,90.25){/}
	\put(54.9,90.25){/}
	\put(96.2,90.25){/}
	\put(10.2,88.75){/}
	\put(51.5,88.75){/}
	\put(54.9,88.75){/}
	\put(96.2,88.75){/}
	\end{overpic}
	\vspace{-95mm}\caption{Comparing the class-conditional dust mass fraction histograms between (a) the traditional and (b) the Wasserstein-regularized CVAE models. The classes encode the input label to the model. Note the non-uniform scale near the top of the vertical axis. The ideal cutoff thresholds between the four categories are denoted with dashed vertical lines at 4\%, 55\%, and 84\%.}
	\label{fig:a03frachists}
\end{figure*}

Figure~\ref{fig:a03frachists} presents class-conditional dust mass fraction histograms for samples generated by the traditional CVAE and by our Wasserstein-regularized CVAE, with colors indicating the input label class used for decoding. Ideally these histograms would not overlap and their cutoffs would align with the training thresholds near 4\%, 55\%, and 84\% dust mass. The traditional model exhibits substantial overlap, including cases such as Figure~\ref{fig:a01tradanec}(a$_3$) where a medium-high dust input label yields a low-dust sample, whereas our method achieves better separation in Figure~\ref{fig:a03frachists}(b). Using Wasserstein Regularization increased the label compliance from $71.9\% \pm 1.5\%$ to $80.2\% \pm 0.9\%$. This improvement is attributable to the additional compliance loss introduced in Section~\ref{sec:042swcvae}.

\begin{figure*}
	\centering
	\includegraphics[page=1,width=0.98\linewidth]{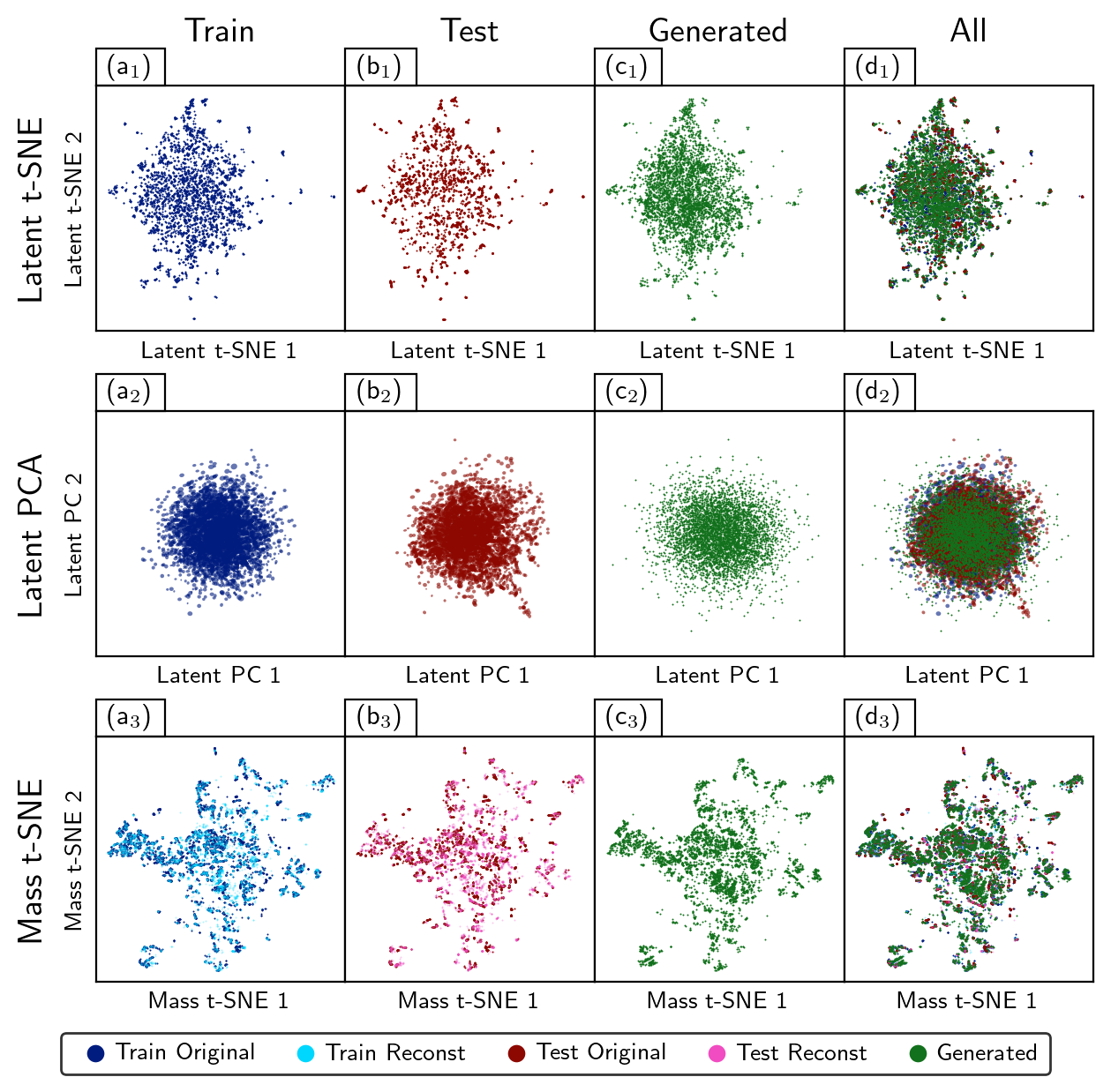}
	\caption{\textbf{Wasserstein-regularized CVAE:} The MDS plots of different data splits.}
	\label{fig:a04wrcvaemds}
\end{figure*}

Figure~\ref{fig:a04wrcvaemds} visualizes split-conditional embeddings for the training, test, and generated data under our Wasserstein-regularized model, showing t-SNE and PCA projections in both the latent and original spaces. With appropriate hyperparameters, the latent representations of train, test, and generated samples occupy the same region, which is reassuring and suggests that generated samples are close in nature to the training and test splits. Nevertheless, potential distribution shift should ultimately be assessed at the decoder output rather than inferred solely from latent-space overlap. Consistent with this, the speciated-mass t-SNE plots indicate that generated samples are reasonably close to the train and test populations.

\begin{figure*}
	\centering
	\includegraphics[width=0.98\linewidth]{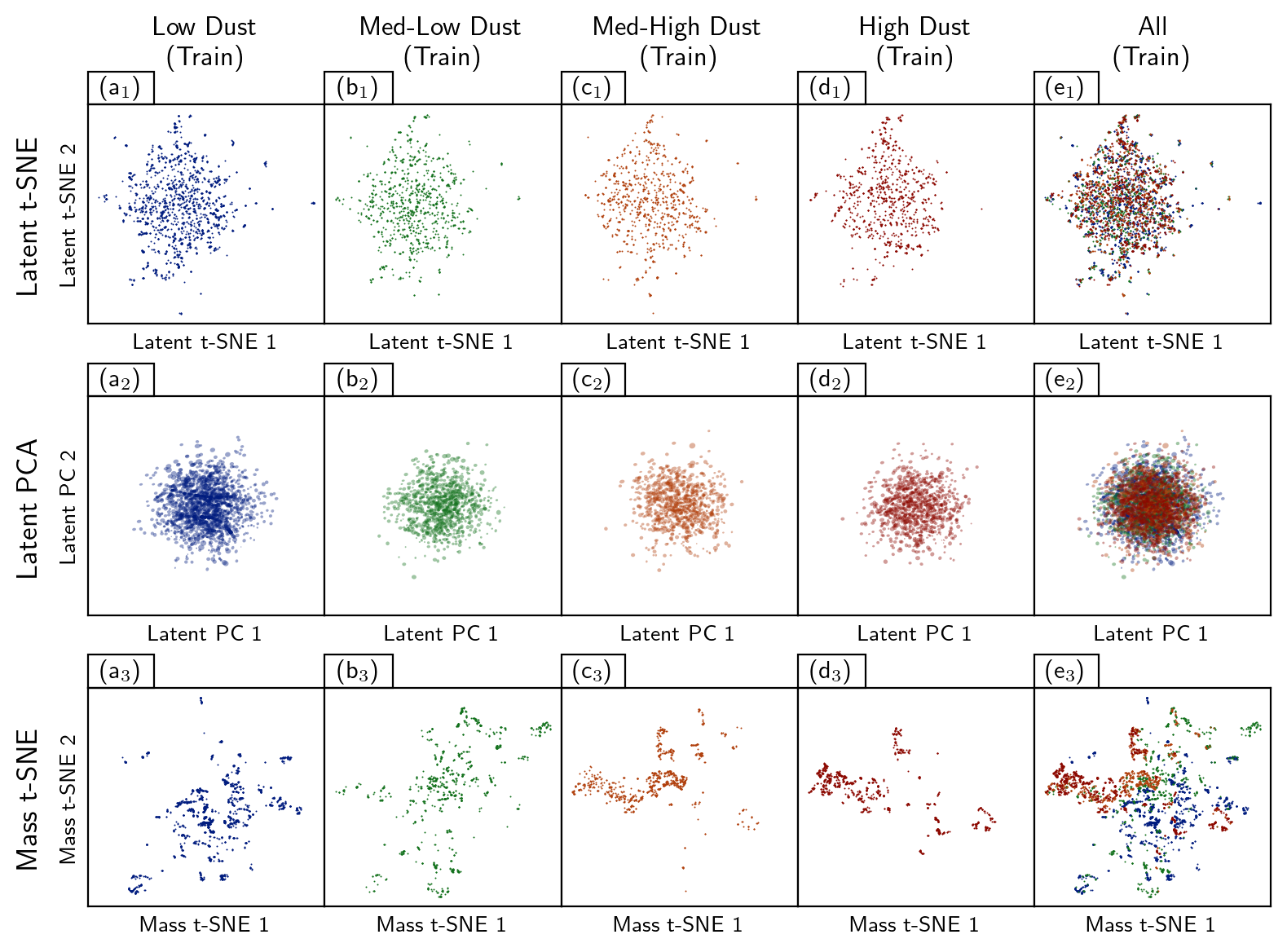}
	\caption{\textbf{Wasserstein-regularized CVAE:} The class-conditional MDS plots of the training data.}
	\label{fig:a05wrcvaemds}
\end{figure*}

\begin{figure*}
	\centering
	\includegraphics[width=0.98\linewidth]{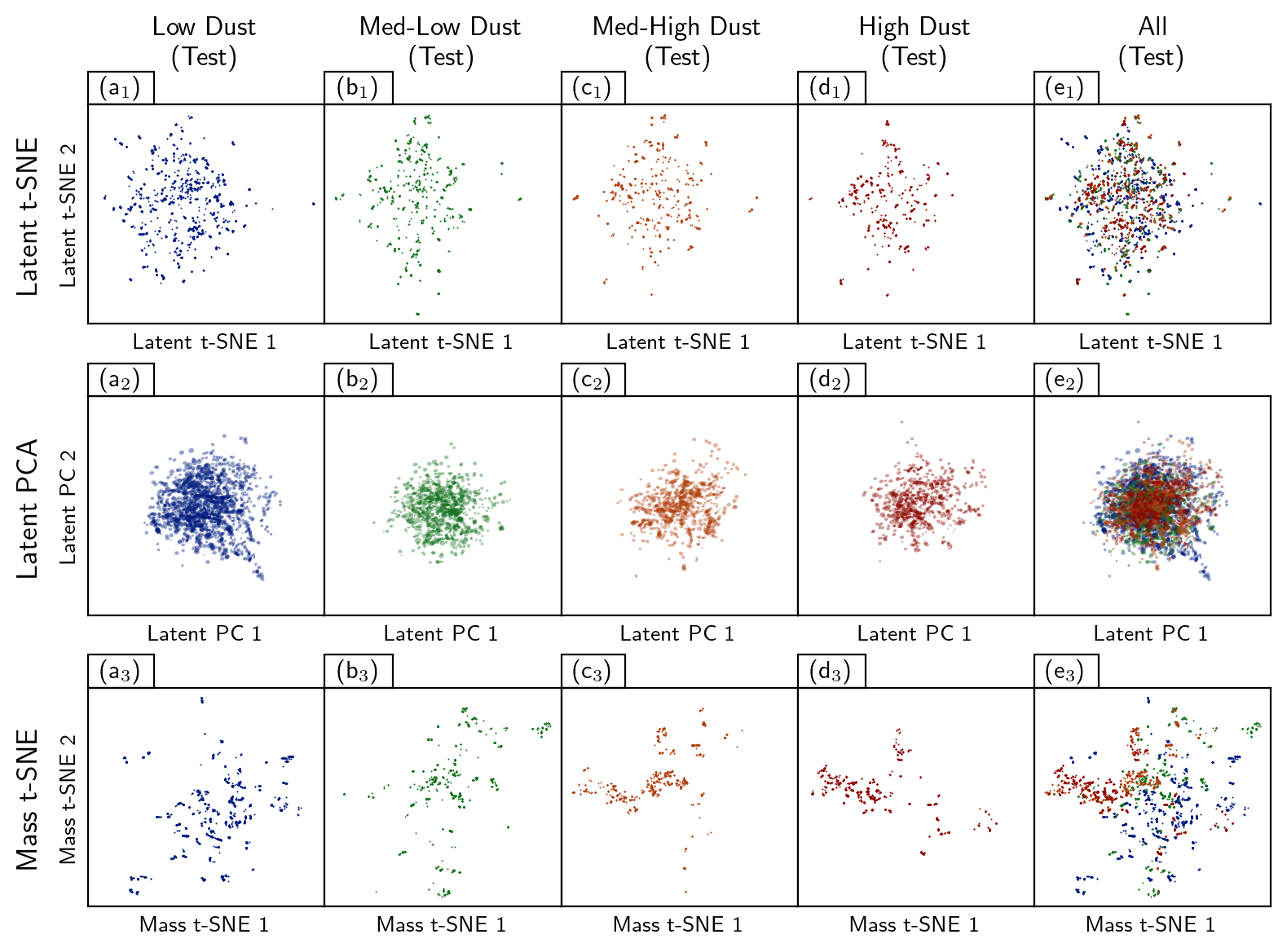}
	\caption{\textbf{Wasserstein-regularized CVAE:} The class-conditional MDS plots of the held-out test data.}
	\label{fig:a06wrcvaemds}
\end{figure*}

\begin{figure*}
	\centering
	\includegraphics[width=0.98\linewidth]{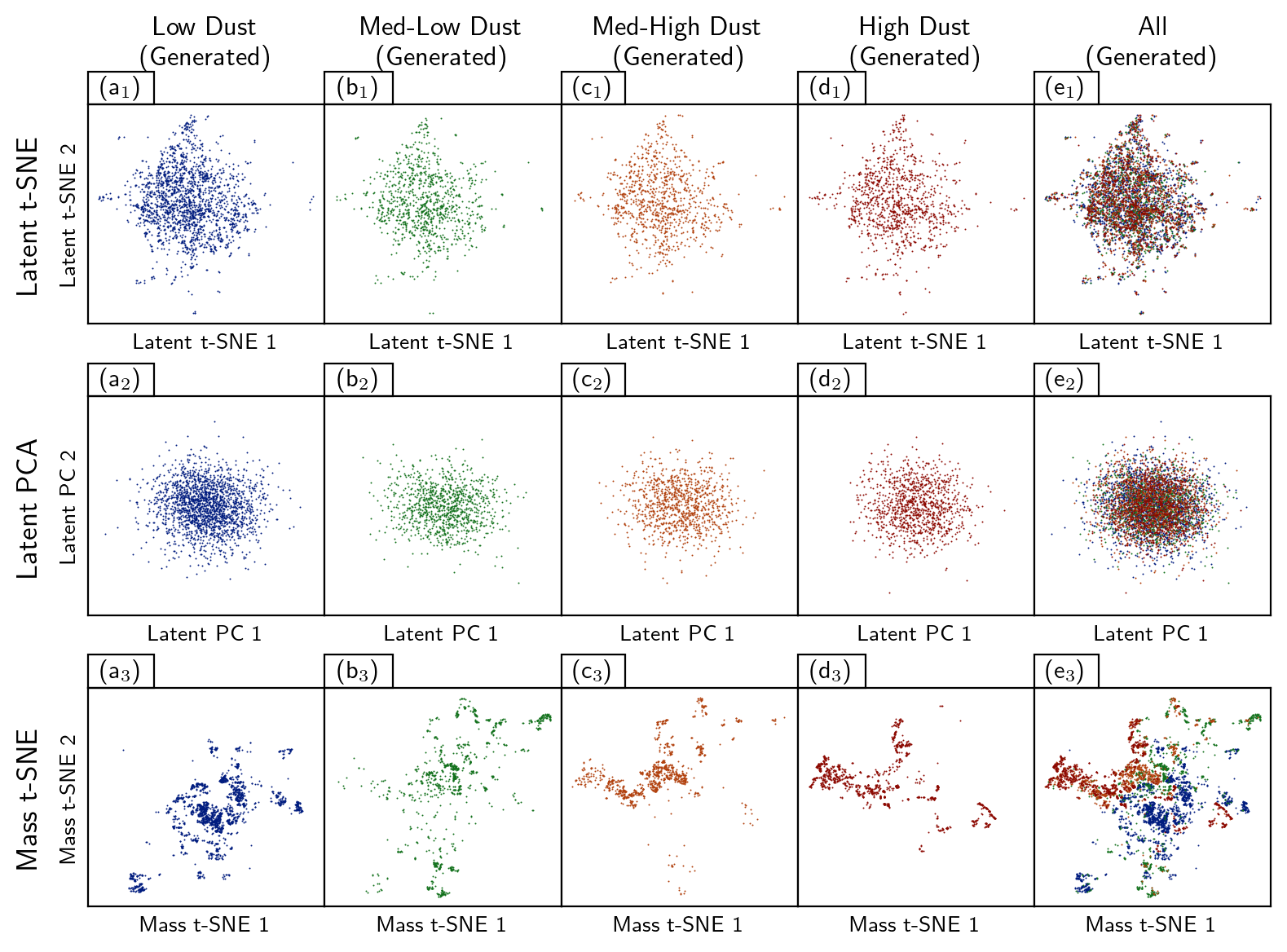}
	\caption{\textbf{Wasserstein-regularized CVAE:} The class-conditional MDS plots of the generated data.}
	\label{fig:a07wrcvaemds}
\end{figure*}

Figures~\ref{fig:a05wrcvaemds},~\ref{fig:a06wrcvaemds}, and~\ref{fig:a07wrcvaemds} show class-conditional MDS representations for the train, test, and generated splits. The latent-space views in Figures~\ref{fig:a05wrcvaemds} and~\ref{fig:a06wrcvaemds} indicate that all categories occupy a similar region, closely resembling the generative latent variables sampled from a normal distribution in Figure~\ref{fig:a07wrcvaemds}; this is desirable because the label and latent variables should be statistically independent and serve distinct roles (Section~\ref{sec:042swcvae}). In the speciated-mass t-SNE plots, the medium-high and high dust categories appear closer and more overlapped than others, consistent with the overlapping histograms in Figure~\ref{fig:a03frachists}, highlighting the difficulty of this setting. While our method improves separation relative to the traditional model (Figure~\ref{fig:a03frachists}), distinguishing medium-high from high dust categories remains challenging.

\subsection{Impact of Wasserstein Regularization: Continuous Labels}\label{sec:a03contlbl}

Tables~\ref{tab:02duogenambig} and~\ref{tab:03trigenambig} compare generative ambiguity for both methods under the low- and high-dimensional label setups. In Table~\ref{tab:02duogenambig}, our method attains the lowest ambiguity in 7 of 9 metrics and is statistically tied in one of the two remaining, with gains varying by diagnostic and measurement and being most pronounced for speciated-mass-related quantities. While our method yields the lowest ambiguity values for all diagnostics and measurements in Table~\ref{tab:02duogenambig}, it is statistically tied in 5 of 9 cases. This reflects the weaker, less-constraining measurement information in the low-dimensional label setup and its similarity to the dust mass fraction scenario in Section~\ref{sec:a02catlbl} of the main paper.

\renewcommand{\arraystretch}{1}
\renewcommand{\ZR}{\phantom{0}}
\begin{table}[h]
	\smaller[0.5] 
	\centering	
	\begin{tabular}{p{0.30\linewidth} >{\centering\arraybackslash}p{0.25\linewidth} >{\centering\arraybackslash}p{0.37\linewidth}}
	\hline
	Generative Ambiguity & \makecell[c]{Traditional CVAE\\Mean [95\% CI]} & \makecell[c]{Wasserstein-Regularized CVAE (Ours)\\Mean [95\% CI]} \\\hline
	Truncated Number Distribution & \textbf{\ZR1.00}\% [\ZR0.95\%, \ZR1.05\%] & \ZR1.33\% [\ZR1.29\%, \ZR1.37\%] \\
	Limited Species Bulk Masses & 12.3\ZR\% [11.8\ZR\%, 12.9\ZR\%] & \textbf{10.8\ZR}\% [10.2\ZR\%, 11.3\ZR\%] \\
	Number Distribution & \textbf{\ZR1.36}\% [\ZR1.21\%, \ZR1.49\%] & \textbf{\ZR1.49}\% [\ZR1.44\%, \ZR1.55\%] \\
	Total Mass Distribution & 45.2\ZR\% [44.2\ZR\%, 46.5\ZR\%] & \textbf{33.9\ZR}\% [33.0\ZR\%, 35.0\ZR\%] \\
	Species Bulk Masses & 50.0\ZR\% [49.1\ZR\%, 51.3\ZR\%] & \textbf{34.9\ZR}\% [33.8\ZR\%, 35.8\ZR\%] \\
	CCN Spectrum & \ZR6.09\% [\ZR6.00\%, \ZR6.21\%] & \textbf{\ZR5.20}\% [\ZR5.11\%, \ZR5.32\%] \\
	Scattering Log-Spectrum & \ZR3.32\% [\ZR3.19\%, \ZR3.45\%] & \textbf{\ZR2.69}\% [\ZR2.59\%, \ZR2.81\%] \\
	Absorption Log-Spectrum & \ZR46.9\% [\ZR46.1\%, \ZR47.8\%] & \textbf{44.2\ZR}\% [\ZR43.5\%, 45.0\ZR\%] \\
	Frozen Fraction Log-Spectrum & \ZR9.36\% [\ZR9.10\%, \ZR9.52\%] & \textbf{\ZR7.87}\% [\ZR7.55\%, \ZR8.14\%] \\
	\hline
	\end{tabular}
	\caption{\textbf{Low-dimensional label setup}: The generative ambiguity values.
	Overall, the generative ambiguity of our method is slightly lower than the traditional method.}
	\label{tab:02duogenambig}
\end{table}

\renewcommand{\arraystretch}{1}
\renewcommand{\ZR}{\phantom{0}}
\begin{table}[h]
	\smaller[0.5] 
	\centering	
	\begin{tabular}{p{0.30\linewidth} >{\centering\arraybackslash}p{0.25\linewidth} >{\centering\arraybackslash}p{0.37\linewidth}}
	\hline
	Generative Ambiguity & \makecell[c]{Traditional CVAE\\Mean [95\% CI]} & \makecell[c]{Wasserstein-Regularized CVAE (Ours)\\Mean [95\% CI]} \\\hline
	Truncated Number Distribution & \textbf{\ZR0.82}\% [\ZR0.71\%, \ZR0.91\%] & \textbf{\ZR0.78}\% [\ZR0.72\%, \ZR0.87\%] \\
	Limited Species Bulk Masses & \textbf{\ZR5.77}\% [\ZR5.10\%, \ZR6.26\%] & \textbf{\ZR5.40}\% [\ZR4.82\%, \ZR6.18\%] \\
	Number Distribution & \textbf{\ZR0.82}\% [\ZR0.71\%, \ZR0.91\%] & \textbf{\ZR0.78}\% [\ZR0.72\%, \ZR0.87\%] \\
	Total Mass Distribution & \textbf{\ZR3.39}\% [\ZR3.18\%, \ZR3.67\%] & \textbf{\ZR3.07}\% [\ZR2.84\%, \ZR3.33\%] \\
	Species Bulk Masses & \ZR5.00\% [\ZR4.78\%, \ZR5.22\%] & \textbf{\ZR3.79}\% [\ZR3.66\%, \ZR4.03\%] \\
	CCN Spectrum & \ZR3.71\% [\ZR3.52\%, \ZR3.83\%] & \textbf{\ZR3.34}\% [\ZR3.18\%, \ZR3.48\%] \\
	Scattering Log-Spectrum & \ZR0.97\% [\ZR0.93\%, \ZR1.02\%] & \textbf{\ZR0.86}\% [\ZR0.83\%, \ZR0.93\%] \\
	Absorption Log-Spectrum & \textbf{17.2\ZR}\% [15.8\ZR\%, 18.5\ZR\%] & \textbf{17.0\ZR}\% [15.9\ZR\%, 18.2\ZR\%] \\
	Frozen Fraction Log-Spectrum & \ZR4.05\% [\ZR3.81\%, \ZR4.36\%] & \textbf{\ZR3.59}\% [\ZR3.41\%, \ZR3.81\%] \\
	\hline
	\end{tabular}
	\caption{\textbf{High-dimensional label setup}: The generative ambiguity values.
	Overall, the generative ambiguity of our method is slightly lower than the traditional method.}
	\label{tab:03trigenambig}
\end{table}


\end{document}